\documentclass[12pt]{article}
\usepackage[margin=2.3cm]{geometry}
\usepackage{setspace}
\usepackage{graphicx}
\usepackage{bm}
\usepackage{mathptm,amsmath} 
\usepackage{graphicx}
\usepackage{amssymb}
\DeclareMathAlphabet{\mathcal}{OMS}{cmsy}{b}{n}

\begin{document}

\doublespacing

\title{{\normalsize \em Running title : Toward a new paradigm for protein sliding}\\
Protein-DNA electrostatics: toward a new paradigm for protein sliding}
\author{Maria Barbi$^a$, Fabien Paillusson$^b$\\
\small  $^a$ Laboratoire de Physique Th\'{e}orique de la Mati\`{e}re Condens\'{e}e, \\
\small Universit{\'{e}} Pierre et Marie Curie, case courrier 121, \\
\small 4, Place Jussieu - 75252 Paris cedex 05, France\\
\small $^b$ Department of Chemistry, University of Cambridge, \\
\small Lensfield Road, CB2 1EW, Cambridge (UK) 
}

\maketitle

\begin{abstract}
Gene expression and regulation rely on an apparently finely tuned set of reactions between some proteins and DNA. Such DNA-binding proteins have to find specific sequences on very long DNA molecules and they mostly do so in absence of any active process. It has been rapidly recognized that to achieve this task these proteins should be efficient at both searching (i.e. sampling fast relevant parts of DNA) and finding (i.e. recognizing the specific site). A two-mode search and variants of it have been suggested since the 70s to explain either a fast search or an efficient recognition. Combining these two properties at a phenomenological level is however more difficult as they appear to have antagonist roles. To overcome this difficulty, one may simply need to drop the dichotomic view inherent to the two-mode search and look more thoroughly at the set of interactions between DNA-binding proteins and a given DNA segment either specific or non-specific. This chapter demonstrates that, in doing so in a very generic way, one may indeed find a potential reconciliation between a fast search and an efficient recognition. Although a lot remains to be done, this could be the time for a change of paradigm.
\end{abstract}


\section{Introduction: the search of target sequences}

The observation of gene regulatory networks made possible by proteomics (the study of the ensemble of proteins in a cell or tissue in given conditions) and transcriptome analysis (the set of messenger RNA resulting from the expression of a portion of the genome of a cell tissue or cell type) reveals the set of interactions between different cellular components. It is then necessary to specify the nature of these interactions, from the structural, energetic, spatial and temporal point of view, in order to reveal the mechanisms underlying the "cellular timing": how appropriate macromolecules are recruited at the right time and at the right place? 

Many proteins indeed have to search and bind specific, relatively short DNA sequences in order to perform their biological task. These specific-proteins include polymerases and a number of transcription factors involved in the regulation of gene expression, but also proteins with different functions as e.g. nucleases. Knowing that the total length of DNA may reach millions or billions of base pairs (bps), one understands that finding the target sequence is a formidable challenge. The problem of this search kinetics have been debated since, in the 70s, researchers realized that the relatively short time needed for a protein to find its target sequence on DNA cannot be explained by a simple search by 3D diffusion in the cell (according to the Smolukowski theory) followed by random collisions with the DNA: the actual association constant is approximately 2 orders of magnitude larger \cite{Rig70,Richter}. Since then, many people have been interested in the search process, and a large amount of theoretical work has been done \cite{von89,ben04}. Interestingly, despite the fact that the role of electrostatics had been explicitly invoked in the original works \cite{Rig70,von89,Hal09}, most of the work has been based on a purely kinetic approach. The main results can be summarized by the finding that 3D excursions should be alternated by phases of 1D diffusion, named {\em sliding}, during which the protein binds DNA and slides along the double helix by thermal 1D diffusion {\em sliding}, during which the protein binds DNA and slides along the double helix by thermal 1D diffusion \cite{Ber81}. This intermittent process has been called ``facilitated diffusion''.

The existence of 1D diffusion or sliding has then been proved by several experiments \cite{Win81,Terry85,Hal04} and in particular by fluorescence microscopy \cite{Shi99,Bla06,Elf07,Bonnet08}. In this kind of experiments the two extremities of a DNA molecule are bound on a surface, in such a manner that the DNA is softly stretched. The movement of a fluorescent protein moving along the DNA direction can then be recorded and analyzed. These experiments confirm that proteins may slide along DNA and generally display a standard diffusion dynamics.
Experiments also show that the sliding lifetime is sensitive to the salt concentration  
\cite{Rig70,Bla06,Bonnet08}. This supports the idea that electrostatics is involved to some extent in the intermittent behavior, with a probable role for the dissolved salt ions.
Electrostatics plays indeed a major role in the protein-DNA interaction  \cite{Jon99,Nad99,Tak92,von07,Via00,Kal04,Carrivain12}. The reason is that DNA is very strongly negatively charged (-2 charges per base pair). On the other hand, DNA-binding proteins are most often positively charged on the surface that faces DNA, so to be attracted onto it \cite{Jon99,Nad99}. 

Together with its role in the search kinetics, the protein sliding is also supposed to have another crucial role: it allows the protein to read the DNA sequence and therefore to distinguish the target site among all other sequences. This reading can be performed, besides other interactions, by the formation of hydrogen bonds between the protein and the side of the base pairs exposed toward the major groove, without opening the double helix \cite{Jon99, Nad99}. Since the patterns of  hydrogen bonds that may be formed on each base pair is different, a protein can discriminate precisely a target site by looking for the formation of the good hydrogen bond pattern along the entire sequence visited.
However (and independently from the precise reading interaction) this reading mechanism leads to a paradox. An efficient discrimination between sequences implies indeed a rough interaction energy strongly varying as a function of the protein position along DNA, and such an energy profile leads in turn to a trapping of the protein, which reduces considerably its mobility. The mobility of the protein seems therefore to be in contradiction with its specificity, i.e. its capability of discriminating the good sequence \cite{Bar04a,Bar04b,Slu04,Slu04b}.
This paradox is not always taken into account in the literature concerning target search, but some authors have addressed the problem. Intuitively, one solution seems to be the existence of two different states for the protein:  one state where the protein slides but cannot recognize the sequence, and another state where it reads but moves in a much slower way. Mirny and co-workers proposed that the protein could undergo conformational changes between a search state and a recognition state, in an intermittent way \cite{Slu04,Slu04b}.   
We have proposed an alternative mechanism, where the key parameter will be the distance between the protein and the DNA \cite{PRL,Pai11}. Since the range of H bonds is rather short, one can guess that this distance can indeed play a crucial role. Our starting point has been the study of the physics of the interaction between protein and DNA, with a focus on the electrostatic interaction. 

A second important ingredient, usually neglected in the modeling of protein-DNA system, emerges from this study: the protein shape.  We have shown indeed that a charged convex body (like DNA) counter-intuitively repels an oppositely charged concave body (like DNA-binding proteins), provided the two bodies do not exactly neutralize each other \cite{PRL,Pai11}. In the following, we will describe how to obtain this result, and discuss its implications on the search mechanism. A possible solution for the {\em mobility-specificity paradox}.

\section{Protein diffusion in the cell}
\label{models}

\subsection{Diffusion: a stochastic regulation tool?}

The search of a target DNA sequence may have a particularly evident biological importance in cell differentiation as evidenced in some recent theories. Among others, JJ Kupiec rejects the predominant role attributed to the "genetic program" (all information necessary for the development of the organism is encoded in the genome) and stereospecificity (for each cellular function there is a specific protein that acts through a deterministic "key-lock" recognition mechanism). An alternative model for cellular functioning is proposed, based on evolutionary approach. In this model, in brief, proteins diffuse into the cell and interact randomly with DNA. Gene expression is also random. However, these interactions are statistically regulated by the position of genes in the cell space and along the genome: the probability of interacting with a closer site is higher, and this effect is strong enough to introduce a differentiation in gene expression. This mechanism finally leads to a kinetic competition that allows to set the appropriate gene expression and to stabilize it as best suited to the needs of the cell \cite{Kupiec}.

Even without adopting this point of view entirely, it is interesting to note that it involves several important elements of the cell functioning. Most of them are nowadays well substantiated. First, it is clear that the affinity of proteins for their target sequences is relative (see e.g. \cite{stormo}). This introduces the problem of obtaining specific recognition while avoiding an excessive competition between slightly different sequences, an effect which can lead to a trapping effect \cite{Ger02, Slu04, Slu04b}.
On the other hand, it is also clear that there is a stochastic component in the search mechanisms, related to the presence of a diffusive dynamics, which allows proteins to move and meet their specific sequences. It follows that gene regulation depends on a stochastic and complex dynamics, and it is therefore appropriate to propose a statistical physics approach to describe regulation, based on a precise description of diffusion, recognition and competition mechanisms.

From the point of view of the diffusion dynamics, target sequences search is indeed a very active research field, involving both theoretical and experimental groups (Halford and Marko wrote a recent comprehensive review of this literature \cite{Hal04}). From the pioneering works of Berg and von Hippel \cite{Ber76,Ber77, Ber78, Ber78b, Win81, von89}, attention has focused on the rate constant of the association reaction between the protein and its target sequence. Then appeared a difficulty: assuming that the protein finds its target by simple random diffusion within the cell leads to reaction times which are too low if compared to those experimentally observed. In 1970, Riggs et al. \cite{Rig70} showed that the association constant of Lac repressor with the initiation site of the lactose operon was two to three orders of magnitude higher than the theoretical prediction of the  Smoluchowski theory for chemical reactions limited by diffusion \cite{Richter}  ($k_a \simeq 10^{10}$ M $^{-1}$ s $^{-1}$ against $ 10^7$ to $10^8$ obtained from the theory). In addition, it was noted that the association constant of the Lac repressor with its specific site is also an increasing function of the length of the flanking {\em non-specific} DNA present in the sample \cite{Win81}. 
This observations suggest the existence of an additional mechanism, involving the interaction of the protein with nonspecific DNA, and able to accelerate the search for the target sequence.

\subsection{3D versus 1D}

It was then proposed  that this particular strategy, able to optimize the target search time and called {\em facilitated diffusion}  \cite{von89}, can be associated with an {\em intermittent} diffusion, composed by several different displacement modes (Figure \ref{fig:hopping}). The newer idea was to include a mode called  {\em sliding} : a one-dimensional, thermal diffusion of the protein along the double helix. The diffusion of the protein during the {\em sliding} has been initially considered either as a free diffusive movement on a (two dimensional) cylindrical surface surrounding DNA, or as a motion along the helical path following one DNA groove. The last hypothesis has the advantage to keep the protein in closest and constant contact with the DNA base-pairs, 
allowing the protein to maintain a specific orientation with respect to the DNA helix.
An helical trajectory has been then indirectly proved for the case of some DNA-binding proteins \cite{Blainey09,Dikic12}, but the question remains open in general \cite{Kampmann04}.

Two other displacement modes, rather similar to each other, are called  {\em hopping} and {\em jumping}, and consist in diffusion excursions in the three dimensional space, allowing the protein to jump to more or less distant sites along the chain. Finally,  during {\em intersegmental transfer} proteins can transiently bind to two  different DNA sites at a time and then directly move from one region to the second one without any intermediate diffusion.

The advantage common to all these mechanisms is to reduce the size of the searched space, thus accelerating the localization of the target sequence. Among them, one-dimensional {\em sliding} has been soon considered as necessary by most authors. The relative weight of {\em sliding}  with respect to three-dimensional diffusion has then be subject to debate \cite{Hal09}. It is obvious that {\em pure} sliding would not be very effective if the starting position of the protein on DNA is far from the target sequence, since the protein will then spend too much time in searching remote regions unnecessarily. This effect is of course enhanced dramatically by the slow  progression that characterizes diffusion (the visited space scales as the square root of time). Under certain assumptions, it is possible to prove that there exists an optimal choice of the mean times spent in 1D and 3D phases respectively,  that minimize the overall target search time \cite{Coppey04}. However, the precise mechanisms governing these two types of motion and the transition from one to the other have still not been elucidated.


\subsection{Experiments: Biochemistry, AFM and fluorescence microscopy}

From the experimental point of view, the possibility to observe one-dimensional diffusion of proteins along DNA has aroused great interest. Biochemical experiments have been performed to measure the average protein-DNA reaction rates as a function of different parameters, and in particular of the lengths of DNA sequences where the target is inserted, were reported \cite{Rig70,Win81,Shi99}. A more quantitative and accurate method, but only applicable to certain proteins, is based on the evaluation of the correlation between the activity levels of a protein in two remote sites located at a known distance  on a DNA molecule ({\em processivity}) \cite{Sta00}. It is interesting to note that, despite its good performances, this experience is open to multiple interpretations \cite{Hal04}, and its results are difficult to reproduce by simple models \cite{Sta00}.
Alternative techniques as atomic force microscopy \cite{Gut99} and fluorescence microscopy \cite{Har99} (Figure \ref{cap}) allow a direct visualization of the protein movement.

The basic principle of the atomic force microscopy (AFM) is to scan the surface of an object by a nanometer sized tip to reconstruct the geometry of the surface. In the case of protein-DNA systems, protein and DNA can either be fixed adsorbed onto the surface, loosely enough to be able to diffuse on it \cite{Gut99}. Despite the very high spatial resolution, this technique was initially limited by a low temporal resolution: tens of seconds between two images. More recently, high-speed AFM allows scanning biological samples in buffer up to 30 frames per second \cite{Ando2001}.

However, another limitation, particularly relevant in the study of diffusion, is due to the presence of the surface itself, which limits the free space around the molecules. Double-stranded DNA immobilized on the
surface may function as a trap reducing Brownian motion \cite{Sanchez}. Similarly, if DNA sliding through  a fixed protein may induce anomalous diffusion as for the passage of a polymer in a pore \cite{Dubbeldam}.

Fluorescence microscopy is used to study processes on large spatial scales
and temporal areas (from nanometer to micrometer and from nanosecond to second) \cite{Bla06,Wan06,Kim07,Bonnet08,Tafvizi08,revue2008}.
The operating principle is simple: the protein is  chemically linked to a fluorescent label (organic fluorophores, fluorescent nano-crystals, fluorescent proteins, quantum dots ...) and can therefore be observed optically. In practice, however, the experience is very sensitive and dependent on many details, particularly related to the properties of fluorescence markers (lifetime of the light emission, flashing...).

Moreover, in order to observe the diffusive motion of a protein around a DNA molecule, it is necessary to fix the DNA in an appropriate manner, in order to immobilize it while leaving the space necessary for the interaction with the protein. Techniques of DNA "combing" have been proposed to this aim. 
Starting from the DNA molecule in its random-coil configuration (the form in which it is found naturally in solution) one of its ends is first bound on a chemically treated glass surface. Then the surface is slowly withdrawn causing the stretching of the molecules by capillarity. Alternatively, combing can be obtained through the application of a hydrodynamic flow of DNA molecules attached at one end: this method enables a more soft stretching, which in addition can be controlled so as to obtain more or less important stretching degrees \cite{Crut03}.

Like any conventional optical microscopy technique, fluorescence microscopy is limited by the diffraction of light. Its resolving power is about 200 nm. However, it is possible to go down to about 30 nm resolution by image analysis techniques for determining the center of the light spot recorded.
This gives a good enough resolution to detect the movement of the protein between two successive images, which are usually separated by a few tens of ms.

An example of the results obtained by fluorescence microscopy is represented by the work of Pierre Desbiolles group \cite{Bonnet08}, an extract of which is given in Figure \ref{cap}. The registration of the position of the
endonuclease EcoRV when bound non-specifically to DNA is decomposed into a longitudinal component and a transverse component. If the latter remains limited, the longitudinal component mean square displacement is proportional to time, consistently with one-dimensional diffusion along DNA. In addition, several dissociation/re-association events are observed, as indicated by a faster movement leading to the re-association on a distant  DNA position in a single time frame, i.e a {\em hopping} process following the usual definition (Fig. \ref{cap}{\bf A} and {\bf B}).

\subsection{Who helps who?}

Thanks to fluorescence microscopy experiments, {\em sliding} has become a reality and its existence as a step in target sequence search is nowadays largely accepted. 
Nonetheless, the actual role of this searching mechanism is still under discussion. 
An important element in this discussion has been the S. E. Halford's paper \cite{Hal09}, where the author contest the need of any mechanism to facilitate the search and affirms that ``no known example of a protein binding to a specific DNA site at a rate above the diffusion limit'' exist.
Indeed, if both 1D and 3D diffusion processes can be observed, the conclusion that facilitated diffusion may greatly enhance DNA-protein association rates is more questionable. 
The point raised by Halford is that the rapidity of these reactions is instead due primarily to electrostatic interactions between oppositely charged molecules \cite{Hal09}.
The large association rates reported in the pioneering work \cite{Rig70} where indeed obtained at very low ionic strength, suggesting a role of the electrostatic attraction that becomes negligible, due to screening effects, in higher salt. This conclusion has however been overlooked in the following literature, until Halford's work.  
We emphasize, in particular, the crucial role attributed to electrostatic, a point to which we will come back in the following. 
 
It is also interesting to note that electrostatic should also determine another important feature of the search process, namely the protein-DNA association strength and therefore the lifetime of the 1D diffusion phase, and therefore the relative weight of 1D and 3D processes. 
This is another important question evoked in discussing the relevance of sliding as a enhancing mechanism in target search. In Ref. \cite{Gow05}, the same S. E. Halford and co-workers showed for the restriction enzyme ecoRV that at low salt, the protein only {\em slides} continuously on DNA for distances shorter than 50 base pairs. Transfers of more than 30  base pairs at in vivo salt, and over distances of more than 50  base pairs at any salt, always included at least one dissociation step. The authors then conclude that 
3D dissociation/reassociation is its main mode of translocation for this protein.

To end this discussion, we would like to point out that that question of the relative role of 3D diffusion and {\em sliding} can also be seen in an opposite way. Due to the electrostatic attraction, indeed, one can take as reference the weakly bound state where the protein stay along DNA. The question is then whether or not {\em 3D excursions} may help the protein {\em 1D} search, and reduce the search time. This is the point of view adopted e.g. by the group of O. B\'enichou \cite{benichoureview12}. 

Whatever the philosophy one adopts, the question of the target sequence search reveals an unexpected richness. Electrostatics seems to be an essential ingredient and, if intermittency is expected to improve the search time in any case, observations and models invoke different {\em sliding} mechanisms (along the helical path or not), together with {\em jumps} and {\em hops}. Moreover, as we will seen in the next section,  alternative {\em sliding modes} have been proposed in order to solve additional difficulties in explaining the protein mobility. It is therefore tempting to ask whether a different ``paradigm'' for the search, based on a different description (or parametrization) of the whole process, may be more adapted.

\section{Diffusion along the DNA: what role for the sequence?}
\label{Diffus}
\subsection{Reading the sequence}

\subsubsection{Direct and indirect interaction}

While experiments on {\em sliding} were multiplying and becoming more refined, this problem was attracting more and more theoreticians, seeking a consistent modeling of the observed phenomena.

Different models have been proposed. However, all models seem to lead to more or less important inconsistencies, and a unified model has not yet been imposed. Some authors \cite{von89, Hal04} consider DNA as a uniform cylindrical space in which the protein is trapped by electrostatic interaction, and could slide spontaneously under the effect of thermal agitation. Some models where the protein would even slide along the helical structure of DNA have been envisioned fairly early \cite{Sch79}. 

However, as some authors stressed rather soon \cite{Bru02,Bar04a,Slu04}, the recognition of the target sequence needs a way of {\em reading} the sequence, which cannot be taken into account by an homogeneous interaction. In order to discriminate the target sequence, it is necessary to introduce a sequence-dependent interaction, albeit small. 

To get a concrete picture of this interaction, let us consider as an example  a particular protein, the RNA-polymerase of T7 virus.
The specific complex formed by the T7 RNA-polymerase and its target sequence (a gene promoter) has been studied by crystallography \cite{Cheetham99} (Figure \ref{fig:T7}). The protein-DNA interaction occurs in three regions: in a first region of 5 base pairs the double helix is bent by the presence of the protein; in a second region, 5 base pairs long, a set of hydrogen bonds between the side chains of the protein and the base pairs is made; finally, in correspondence of a third site, a portion of the protein in inserted between the two helices of DNA causing a local opening of the double helix.

Among the different interactions, some are likely to participate in the target sequence search, others are probably induced only once the target is reached.
The latter interaction, which characterizes the formation of the {\em open complex}  (the pre-activated state, ready to start the gene transcription), is most probably absent during the search. The two other modes of interaction are two typical example of direct (chemical) and indirect (mechanical) interaction \cite{Pai04}.
The first interaction will include, typically, direct hydrogen bonds to base pairs  and Van der Waals interactions \cite{See76,Nad99}. 
Hydrogen bonds provide the higher level of sequence specificity, and may be used to define a simple code to explain sequence reading. In the following, we will precise how this specificity is obtained.

Entropic contributions due either to the loss of degrees of freedom of the protein and DNA, or to the expulsion of ions and water molecules from the protein-DNA interface, may also contribute to the direct par of the interaction, but their degree of specificity is less easily quantified. 
 
On the other hand, sequence-dependent changes in DNA structure, or in its mechanical or dynamic properties, can also play a role in recognition \cite{Pai04}. Sequence-induced protein deformations may also be considered.
Such mechanical effects may be used by the protein as discriminating tools.They may give rise to rather smooth energy profiles \cite{Slu04}, correlated over distances comparable to the length of the target sequence (Figure \ref{fig:slusky2}), and has interesting dynamic properties not yet fully explored.

\subsubsection{Hydrogen bonding}

Let us now just take into account the hydrogen bond contribution to the overall interaction, and precise its origin.
All DNA base pairs expose in the major groove a regular pattern of four chemical groups that can be donors or acceptors of hydrogen bonds (Figure \ref{fig:Hbonds}). On the other side, a protein like the T7 RNA-polymerase presents a reactive site that contains, through the arrangement of its side chains, a recognition pattern containing the information on the correct disposition of donor and acceptor groups in the target.
It seems reasonable to assume that the protein looks for this same pattern on any sequence during the search. We also assume that the H-bonds formed
in the DNA-protein complex at the recognition site are known (this information
can be obtained from crystallographic analysis of the DNA-protein complex).
The interaction between the protein and a given sequence can therefore been simply described by counting the number of bonds it can make at that position, i.e. the number of DNA groups that are consistent with the protein recognition pattern. 
Within this model, the protein can be represented by a \emph{recognition
matrix} containing the pattern of H-bonds formed by the protein and the
DNA at the recognition site.

When the protein is at position $n$, the sequence that it is visiting can be represented as a list of vectors, $D^{(n)}  = b_{n+1}, b_{n+2}... b_{n+N}$, where
\[
b_{n}=\left\{
\begin{array}{l}
(1,-1,1,0)^{T}  \,\,\,\mbox{for base A}\\
(0,1,-1,1)^{T}  \,\,\,\mbox{for base T}\\
(1,1,-1,0)^{T}  \,\,\,\mbox{for base G}\\
(0,-1,1,1)^{T}  \,\,\,\mbox{for base C}
\end{array}
\right.\] 
and where the number $N$ of vectors correspond to the length of the visited sequence. 
The recognition matrix is then a N$\times$4 matrix containing the ``good'' pattern of hydrogen bonds, i.e. the one that will be made on the target.
In the specific case of the T7 DNA-polymerase, e.g., the recognition matrix reads
\begin{equation}
R = \left(
\begin{array}{rrrr}
1 & 1 & 0 & 0 \\
1 & -1 & 0 & 0 \\
1 & 1 & 0 & 0 \\
0 & 1/2 & 0 & 0 \\
0 & 0 & 1/2 & 1
\end{array}\right)
\end{equation}
where the factors $1/2$ have been introduced in order to reproduce one
hydrogen bond shared by two base pairs.
The interaction energy for the protein at position $n$ is then 
given by the sum of all positive matches and can be written as
\begin{equation}
  \label{eq:energy}
  E(n)= \, - \, {\mathcal E} \sum_{i=1}^N \sum_{j=1}^4 max(R_{ij} D^{(n)}_{j}, 0)\,,
\end{equation}
where $- {\mathcal E}$ is the net energy gain of a single hydrogen bond (of the order of a fraction of $k_B T$  \cite{Tar02}, see discussion below).

Equilibrium measurements \cite{stormo} reveal that the binding energy of a protein to a given sequence can be described, to a good approximation, as the sum of the binding energies to the single base pairs composing the sequence. 
If the latter can be assumed as independent, then the binding energy can be reasonably described as a Gaussian random variable \cite{Ger02,benichoureview12}.  This is indeed what is measured for some real cases \cite{benichoureview12}; in the case of the T7 RNA-polymerase, the same results has been derived based on a detailed analysis of the protein hydrogen bond pattern \cite{Bar04b}.

\subsection{The recognition-mobility paradox}

The one-dimensional diffusion along DNA ({\em sliding}), apparently simple, may hide an unexpected complexity. Most of the authors assume however for this diffusive phase a  simple diffusive dynamics or {\em normal diffusion}. In this case, the mean square distance traveled by the protein along DNA after a time $t$ is proportional to time, i.e.  $ \langle r^2 \rangle = 2D\,t $, where the only parameter that remains to be fixed is the diffusion constant $D$.
Now, if this model is appropriate when the interaction energy is absolutely uniform along the DNA, it is no longer valid when a sequence dependent energy profile is taken into account.

Starting from the previous definition of the protein-DNA 1D energy profile\footnote{The energy profile described here may  be enriched by adding energy barriers for the translocation from any DNA position to the next one. The results are quantitatively, but not qualitatively, affected.}, it is easy to model the one-dimensional diffusion. 
The protein moves  by one-site steps on the energy
landscape $E(n)$, with rates  of translocation between neighboring sites $n$
and $n' = n \pm 1$ defined according to the {\em Arrhenius law}, i.e. proportional to
$\exp{(- \beta\,\,  \big( E(n')-E(n) \big) )}$ whenever $ E(n')-E(n)>0$, while it is constant if $E(n')-E(n) \le 0$. Both expression can be formally written as an identical exponential term of the form $\exp{(- \beta\,\, \Delta E_{n \to n'})}$ by defining $\Delta E(n \to n') = {\rm min}\big(E(n')-E(n), 0  \big)$.
If, moreover, we want to include
a nonzero probability for the protein to stop at one position, the
complete set of translocation rates will reads:
:
\begin{eqnarray}
  \label{eq:rate}
r_{n\to n'} 
&=& 1/2\, \exp{(- \beta\,\, \Delta E_{n \to n'})}, \,\,\,\,\,\, n' = n \pm 1 
\nonumber \\
r_{n\to n} 
&=& 1 -  r_{n\to n+1} - r_{n\to n-1}  \,,
\end{eqnarray}
where $\beta=1/k_B T$. 

Note that the case $\Delta E_{n \to n'}=0$  $\forall n$ corresponds to a constant energy landscape, i.e. to a simple 1D diffusion process with diffusion constant $2D = 1$. This limit can also be recovered in the case where ${\mathcal E} =0$.

The numerical study of this diffusion process gives a predictable result: the trapping effect due to the roughness of the potential gives rise to subdiffusion \cite{Bar04a,Bar04b}.  Figure \ref{fig:subdiff} show this effect as a function of the potential roughness $\beta {\mathcal E}$. 

In the limit of a $\beta { {\mathcal E}} = 0$, i.e. in the case of a flat
underlying potential, the diffusion is of course standard,
with $D=1/2$ and a linear dependence on time, so that the
corresponding curve is a straight line of slope 1 in the log-log plot.
For larger values of $\beta  {\mathcal E}$, the dynamics shows initially
large deviations from the normal diffusion: in these finite
temperature cases,the mean square distance is no longer proportional to time, but increases as a power of time which is smaller than unity, according to the law 
\begin{equation}
 \langle \Delta n ^2\rangle = A\, t^b\,, \,\,\,b<1 \,.
\end{equation}

This effect is transitory: the diffusion becomes normal when one considers  long enough time. Accordingly, 
the exponent $b$ increases  with time toward its equilibrium value of $1$.  This is due to the characteristics of the energy profile, which is rough, but bounded. Roughness thus affects the diffusion for short times, i.e relatively small distances, but it is smoothed out when longer displacements are considered. Overall, on long time scales it only affects the average. 

However, the lifetime of the non-specific DNA/protein complex can be relatively short: normal diffusion behavior can never be reached, and subdiffusion may be the most appropriate description of the protein motion. Moreover, even if the normal diffusion regime is reached, the transitory sub-diffusive phase will significantly change the overall distance travelled by the protein after a given time. 
By focussing for instance on the time needed to perform a mean squared displacement of 100 bp$^2$ (therefore a typical distance of 10 base pairs), we can see from  Figure \ref{fig:subdiff}, we can see that this time can be increased by  up to three order of magnitude for the values of ${\mathcal E}$ used. 

In conclusion, this deviation from normal diffusion is not a  merely academic question: all quantitative estimates made to determine the respective roles of 1D and 3D search would be affected and should be recalculated in view of these results.
We stress that it is not easy to obtain a reasonable estimate of the ${\mathcal E}$ parameter. However, rough estimates based on typical hydrogen bond energies (of the order of a fraction of $k_B T$  \cite{Tar02}) do not seem compatible with the double requirement of a protein which has to be free enough to slide along the DNA molecule but also able to bind its target sequence with an energy much higher than for other sequences, so as to ensure a good specificity \cite{Bar04b, Slu04, Slu04b,benichoureview12}. 
In this sense, the trapping effect observed in this simple model evidences the existence of a {\em recognition-mobility} paradox  (also called {\em speed-stability} paradox in the literature). A different way of presenting the paradox, although leading to the same conclusions, is to show that disorder in the binding energy profile on which diffusion takes place leads to 
an effective diffusion constant that decreases exponentially with the variance of the energy distribution \cite{Zwa88,Slu04b,benichoureview12}. Essentially, the requirement of a reasonable specificity prohibits the protein to diffuse. 

\subsection{Two-state models}

To solve this paradox, some new mechanisms have been invoked. One of them can be a modified energy distribution where the binding energy at the target is reduced without affecting the energy distribution variance. However, experimental data does not support this hypothesis \cite{benichoureview12}. 

An alternative solution may be associated with protein conformational fluctuations, this leading to introduce ``two-state'' models. In brief, the idea is to provide two different 1D {\em sliding modes}: a first, {\em reading mode}, where the protein is able to {\em read} the sequence with a reduced mobility, and a second, {\em diffusing mode}  where the protein is able to move relatively rapidly along the double helix, but is essentially blind to the sequence \cite{Win81,Mirny09,Murugan10,Zhou11}.
The conformation change was initially attributed to a microscopic binding of the protein to the DNA accompanied by water and ion extrusion, but such a transition is usually accompanied by a large heat capacity change \cite{Spo94} that in turn
need significant structural changes to be accounted for. Hence, it has been proposed that these two states can be associated to distinct conformational states of the protein-DNA complex \cite{Ger02}, eventually associated to a partial protein unfolding (in the  {\em diffusing mode}) \cite{Slu04b} (Figure \ref{fig:Slusky}). 
However, this mechanism is only efficient if an effective correlation between  the transitions between the two modes and the ``underlying'' energy profile exists. In this way,  the transition  to the {\em reading mode} happens mainly when the protein is trapped at a low-energy site of the search landscape, this being related to a mechanism based on residence times \cite{Slu04b}.

A recent analysis of the efficiency of such mechanism seems to rule out these models, based on quantitative estimates of the relevant parameters \cite{benichoureview12}.
Similarly, it is shown that the presence of a large number of copies of the same protein can resolve the  {\em recognition-mobility} paradox only if the 
energy profile has a small variance \cite{benichoureview12}.
Instead, a new mechanism which is based on {\em barrier discrimination} is proposed, 
which allows to obtain a possible solution fir the process  \cite{benichoureview12}. The basic idea is again that the protein has two different conformations, but the additional element is that these conformations are separated by a free-energy barrier
whose heigh {\em depends} on the position along DNA.
This implies a differences between transition rates from the {\em diffusing} to the
{\em reading} mode that finally allow the protein to improve it search time  as requested.

But how can this model be justified from a physical point of view? A rationale for this model had already been proposed, based on a more  {\em physical}  approach to DNA-protein interaction \cite{AZBEL,Pai09} : we will develop it in next section.

\section{Electrostatics. The DNA-protein interaction}

\subsection{DNA}

In the approaches to the study of the kinetics of protein search described until now, the physics of the DNA-protein interaction is only indirectly taken into account. In particular, a description of the {\em electrostatic} interaction between the two macromolecules in solution was completely missing. In reality, as already mentioned, electrostatics plays a fundamental role in this system. 

The mechanical behavior of a DNA molecule of given length can be described, in an effective manner, by different models of polymers \cite{Cocco}.
Different models can be in rather good agreement with experimental results for force-extension experiences, typically performed using  optical or magnetic tweezers. In this set up, one end of a DNA molecule is bound to a flat substrate, and the other end to a colloidal bead that can be manipulated by an external optical or magnetic field, so to exert a force on the bead and thus on the DNA molecule. The best fit of the resulting data is given by the
{\em Worm Like Chain} model, describing the DNA as an elastic rod (Figure \ref{fig3}). The torsional rigidity of the rod is accounted for by a given value of the {\em persistence length} $L_p $ \footnote{Explicitly, the persistence length can be defined as the characteristic length of the exponential decreasing of the angular correlation of the tangent vector to the polymer (see e.g. \cite {Cocco}).}. For DNA,  $L_p $ is about 50 nm, i.e. approximately 150 base pairs. This is a  quite unusual value for a polymer of  $\sim$  2 nm thickness:  we could expect a higher flexibility at a scale much larger than the thickness. 

This large persistence length depends on an aspect of DNA that have not yet discussed: it is a polyelectrolyte, i.e. a charged polymer. Each phosphate group in the  DNA backbone is indeed negatively charged. Since there are two phosphate groups per base pair in double-stranded DNA, this corresponds to a linear charge density of the order of -2$e$ per base pair (3.4 nm), or -6$e$/nm, or finally a surface charge density of the order of -1$e$/nm$^2$. In comparison, if a power cable in the air had the same surface charge density, the potential difference with respect to the ground would be four orders of magnitude larger than the breakdown voltage in dry air. The DNA molecule is therefore a highly charged molecule. As a consequence, the phosphate groups strongly repel each other, despite the screening effect due to ions in solution. This adds to the natural rigidity of DNA an additional stiffness, that justifies its large persistence length. 
At the protein scale, which is of the order of a few tens of nanometers, the DNA molecule can therefore be modeled as a rigid cylinder of radius $R_ {\rm DNA} = $1  nm, carrying a constant surface charge density of  -1 $e$/nm$^2$. 
For simplicity, we can also assume that the dielectric properties of DNA are those of pure water, i.e. $\varepsilon_{\rm DNA} = \varepsilon_w = $ 80.

\subsection{Proteins}

\subsubsection{Charge}

Non-specific interactions between proteins and DNA are poorly documented, but the predominance of electrostatic undeniable \cite{Jon99, Nad99, Tak92, von07, Via00, Kal04}.
Proteins that bind to DNA are most often positively charged. More precisely, 
positively charged {\em patches} are observed in the region which faces the DNA when the specific complex is formed, an effect which can be accounted for by evaluating the {\em propensity} of positive residues to occurs more frequently in a DNA-binding interface  \cite{Jon99, Nad99, Stawiski03, Jon03, Ahmad04,Szilagyi06, PRL}.

As an illustration of this effect, we show in Figure \ref{fig:history} an analysis of 
the large dataset of DNA-binding proteins features presented in Ref. \cite{Jon99}.
Among the proteins analyzed in this work, it is possible to identify a large family of specific proteins, i.e. binding to specific sequences: this family includes transcription factors, TATA-binding proteins, and restriction enzymes. Other non-specific proteins such as eukaryotic polymerases, repair proteins, histones, form a second group.
We evaluated the surface charge of these proteins in the region of interaction with DNA by counting the charged residues at the interface, and we obtained a very interesting histogram of the charge densities.
In all cases, the DNA-protein interface results to be positively charged. Interestingly,  in the case of proteins that recognize specific sequences, such as transcription factors and restriction enzymes, we obtained an average density of surface charge $ \sigma_{\rm prot} = (0.17 \pm 0.03) e$ nm$^{-2}$.
Besides, we find that non specific proteins are more charged: we get $ \sigma_{\rm prot} = (0.27 \pm 0.05) e $ nm$^{-2}$.

Now, the main role of the positive charge of the protein is, of course, to create an electrostatic attraction to DNA. But the difference observed between different classes of proteins, and the fact that their charge seems to be rather finely tuned, suggest that the surface charge may have a more precise  function in the interaction with DNA, that it would be interesting to elucidate.

\subsubsection{Shape}

If the charge of the protein immediately appears as one of the main ingredients in an electrostatic model of the protein-DNA interaction, another potentially essential ingredient is less easily recognized. Yet, one of the most characteristic aspects of the DNA-binding proteins is their shape complementarity with DNA. DNA-binding proteins often have a concave shape that fits closely DNA. They can cover the DNA molecule by using up to 35 \% of their surface \cite{Jon99}. 
Averaging over different types of proteins, one obtains for the average surface of the interface a value of $ S_{\rm prot} \sim 15 $ nm$^2$ \cite{Jon99,Nad99,Stawiski03,PRL}. 
Generally, and particularly for enzymes, electrostatic patches and significant protein concavities often overlap, so that DNA is "inserted" in this concavities leading to a quite typical {\em enveloping} or {\em complementary} shape \cite{Jon99,Stawiski03} (Figure \ref{fig:jones}).

This shape complementarity of DNA-binding proteins and  DNA enables to maximize the number of direct interactions with DNA base pairs \cite{Jon99,Nad99}. 
Interfaces of DNA-binding proteins have indeed on average more potential hydrogen bonding groups (more than twice as many)  compared to regions that do not bind DNA  \cite{Stawiski03}.
In the specific complex, these bonds may the protein closely stack to DNA, so that interfaces exclude solvent molecules from the interstitial space. However, it is tempting to ask whether this particular protein shape may play a role in  {\em non specific} protein-DNA interactions, at work during the target sequence search.
In this regard, it is interesting to note that structural studies of some non-specific protein-DNA complexes show a gap between the two macromolecules, filled with solvent \cite{Jon99, Nad99, von07, Via00, Kal04}. This observation suggests the existence of a force that counteracts the electrostatic attraction. If this is the case, the question arises as to the physical origin of this repulsive force, and how it depends on the precise value of the surface charge of the protein.

\subsection{A Monte Carlo study}

In order to describe the electrostatic interaction between protein and DNA and the role of the protein charge and shape, we developed a minimal model of DNA-protein system to be studied by Monte Carlo simulation \cite{PRL, Pai11}. We modeled the DNA as a regular cylinder, two nanometers in diameter. To compare different protein shapes, we modeled the protein by simple solid bodies: either a sphere, or a cylinder, or a cylinder with a cylindrical cavity. Hollow cubic shapes have been also tested. DNA charges are placed on its axis, protein charged are placed just below the surface which faces the DNA. The relative orientation between the protein and the DNA was fixed so to orient the charged surface of the protein toward DNA. The distance $L$ between two objects was then varied.

The two bodies are placed in a simulation box with periodic boundary conditions, where water and ions are described by {\em primitive model} \cite{Han00}: the solvent is treated as a continuum dielectric medium with dielectric constant $\varepsilon_w$,  while all ions are modeled by small charged spheres of radius $0.15$ nm. Monte Carlo simulation was done in the presence of monovalent salt corresponding to physiological conditions (0.1 mol L$^{-1}$, or 0.06 molecules nm$^{-3}$).The electrostatic forces acting between protein and DNA  can then be calculated, and integrated to obtain the free energy profile as a function of the DNA-protein distance $L$  \cite{DahirelPRE07,DahirelJCP07}.

Monte Carlo simulations show that while the overall shape of the protein has little influence on the interaction, its complementary with DNA is crucial. 
The complete comparison of the different protein models have been presented in Ref. \cite{PRL}.
The main result of this analysis is presented in Figure \ref{fig1}, where the free energy profiles obtained with the spherical and complementary shapes shown in Figure \ref{fig:shape} \cite{PRL,Pai11}.
 While in the case of a spherical protein the electrostatic interaction is always attractive, in the case of complementary surfaces a repulsion appears below a distance $L$ of a fraction of nanometer (0.1 to 0.75 nm, as a function of the protein charge). A "naive" modeling of the protein as a sphere might be therefore not suitable for the study of the electrostatic interaction!
This result is remarkable: above a distance of the order of a nanometer, the protein is {\em repelled} instead of being attracted by DNA. We will discuss the possible biological role of such an effect in Section \ref{paradoxsolved}, but before, we would like to give a closer look at the physical mechanism leading to this rather surprising effect.

\section{Theoretical approach}

What is the physical origin of the repulsion? It is obviously related to the fact that the two charged bodies are immersed in an ionic solution: the physical description of the system will therefore require some notion from colloidal systems physics. On the other hand, Monte Carlo simulations showed that this repulsion is related to the presence of the two complementary surfaces, which create a large interface between the two charged macromolecules. We can then assume that for small distances between the two bodies, the system can be reasonably approximated by two planar charged surfaces approaching one another (e.g. the DNA plate at $x=0$ and the protein one at $x=L$, as in Figure \ref{fig:twoplates}). This model is very simplified but, precisely for this reason, can be solved by a semi-analytical approach \cite{Parsegian72, Ohshima75, Lau, Safran07, Pai09} whose physical insights are summarized in this section.
We will see that having monovalent ions in solution has two consequences on the attraction between two oppositely charged plates. First, ions generate an osmotic repulsion, due to the loss of available space for them to move as the plate-to-plate distance decreases. Second, a screening effect due to the presence of a salt in solution. To gain as much physical insight as possible we shall introduce these two aspects one at a time, starting with the osmotic repulsion.

\subsection{Counterions only}

We start considering a protein-DNA system modeled as two plates with only one type of monovalent counterions in between so as to ensure electroneutrality (Figure \ref{fig:twoplates} {\em (b)}). On the one hand, if $\sigma_{\rm DNA} <0$ and $0< \sigma_{\rm prot} < |\sigma_{\rm DNA}| $ respectively denote DNA's and protein's surface charge densities, then the {\em direct} electrostatic force per unit area between them is $\Pi_{\rm elec} \approx -|\sigma_{\rm DNA} \sigma_{\rm prot}|/2 \epsilon$. On the other hand, modeling the ions as an ideal gas in a slit of width $L$, the corresponding osmotic pressure is $\Pi_{\rm osm} \approx n_c k_B T$ with $n_c = (|\sigma_{\rm DNA}|-\sigma_{\rm prot})/L$. Balancing these two pressures yields an equilibrium distance that reads:
\begin{equation}
 L_{eq} = |\lambda_{\rm DNA} - \lambda_{\rm prot}| \label{counterions}
\end{equation}
where we introduced the Gouy-Chapman (GC) length $\lambda_X = 1/(2\pi l_B |\sigma_X|)$ for a plate with surface charge $\sigma_X$ (in units of $e$ per unit area) and where $l_B = e^2/(4\pi \varepsilon k_B T)$ is the Bjerrum length. In this first limiting case, we have therefore easily estimate the equilibrium distance between the two plates, due to the imbalance between electrostatic attraction and ion osmotic pressure.

A comment on GC length will be useful. The GC length represents the width of the layer of counterions condensed at the plate of charge $\sigma_X$ they neutralize. It can be retrieved by seeking at what distance from the plate a condensed counterion would go because of a thermal fluctuation. 
The counterion density at a distance $x > 0$ from the charged plate\footnote{The given formula works when one considers a plate and a fully neutralizing solution on its right i.e. there is no electrolyte on the left of the plate.} reads $n_c(x) = (\lambda_X + x)^{-2}/(2\pi l_B)$ \cite{Lau}. Two things are worth noting from this formula. Firstly, the density at zero is $n_c(0) = \sigma_X/\lambda_X$. This result is easily understandable from a physics point of view, since it could have been obtained  by imagining that all the counterions are trapped in a layer of width $\lambda_X$. Secondly, since the charge density is not uniform and actually decays as $x$ increases, the cumulative ionic charge over $n$ GC lengths is $\sigma_X(1-1/(n+1))$ so that for $n=1$, only 50 \% of the charge of the plate is screened (instead of the 100\% one would have guessed from the density at the plate and with uniform assumption).

\subsection{High salt concentration}

When salt with bulk concentration $n_b$ is added to the system, each counterion has a screened electrostatic interaction with the others and, at a coarser level, the plates also have a screened electrostatic interaction. This screening effect is accounted for by a unique parameter called the Debye screening parameter, $\kappa \equiv \sqrt{8\pi l_B n_b} $ for a $1:1$ symmetric electrolyte. It is more intuitive to look at the inverse Debye parameter, $\lambda_D = \kappa^{-1}$, called the Debye length,  that can be understood as the effective range of the electrostatic interactions in solution. 

The osmotic effect, still related with ions thermal motion, plays two different roles when salt is added. First, trapped counterions tend to repel the plates; second, bulk ions tend to increase their accessible volume at the expense of the volume between the plates,  and therefore contribute attractively to the osmotic pressure.  
At high salt concentration,  the resulting positive excess osmotic pressure in between the plates reads \cite{Parsegian72} $\delta \Pi_{\rm osm} = 2n_b(\cosh \psi -1)k_B T \approx n_b \psi^2 k_B T $ where $\psi(x)=\beta e \phi(x)$ is the dimensionless electrostatic potential at $x$. 
If we moreover imagine that at close protein-DNA distances $L$ the dimensionless potential is dominated by the most charged plate (i.e. the DNA plate), then we have at the protein plate $\psi \approx 2\lambda_D e^{-\kappa L}/\lambda_{\rm DNA}$ and $\delta \Pi_{\rm osm} \approx 4 n_b \lambda_D^2 e^{-2\kappa L}/\lambda_{\rm DNA}^2$. 

Since the electrostatic force is screened,  we can assume that at the protein plate it equals  $\Pi_{\rm elec} \approx -|\sigma_{\rm DNA} \sigma_{\rm prot}|e^{-\kappa L}/2$. As before, equating these two contributions allows one to get an equilibrium distance:
\begin{equation}
 L_{eq} \approx \lambda_D |\ln \frac{\lambda_{\rm prot}}{\lambda_{\rm DNA}}| \,.
 \label{salt1}
\end{equation}

Although the assumptions we used to derive Eq. \eqref{salt1} in a simple manner seem very restrictive, this last result is much more robust and holds whenever the salt concentration is high \cite{Parsegian72,Ohshima75,Pai09}. It is also worth noting that Eq. \eqref{salt1} can be rewritten in a way similar to Eq. \eqref{counterions} by introducing an effective counterion cloud size at high salt concentration $\lambda_X^{\rm salt} \approx \lambda_D(\ln 2 + \ln \kappa \lambda_X)$ so that Eq. \eqref{salt1} reads now:
\begin{equation}
 L_{eq} \approx |\lambda^{\rm salt}_{\rm DNA}-\lambda^{\rm salt}_{\rm prot}| 
 \label{salt2}
\end{equation} 

The expression given for $\lambda_X^{\rm salt}$ cannot be interpreted as simply as the GC length because the presence of salt in the system imposes one to choose explicitly a gauge for the potential $\psi$ \cite{Tamashiro03}. In practice, the potential offset is commonly chosen so as to be zero in bulk solution (i.e. far away from the plates). This implies that in a high salt regime the potential $|\psi_0|$ at the plate is of order $\mathcal{O}(1/(\kappa \lambda_X)) \ll 1$ and asking at which distance from the plate a fluctuation $k_BT$ can bring a counterion does not make sense in this context (while it did in absence of salt). Finding an interpretation is not desperate however and one can check easily that at a distance $n\, \lambda_X^{\rm salt}$ away from the plate, the potential is of order $\mathcal{O}(1/(\kappa \lambda_X)^{n+1}) \ll |\psi_0| \ll \mathcal{O}(1)$. Hence, each step $\lambda_X^{\rm salt}$ away from the plate decreases drastically --- by the same proportion --- the potential toward zero. Another way to look at this question is to compute the cumulative charge over a width $n \lambda_X^{\rm salt}$ from the plate. This quantity scales as $ \sigma(1-1/(2\kappa \lambda_X)^n)$: hence, almost 100\% of the plate charge is screened by this cumulative charge and we now exactly how far it is from 100\%. Finally, note that the cumulative ionic charge in the high salt case is much faster closer to the charge plate $\sigma_X$ than in the counterion case. This reflects the very different behavior of the charge density in those two cases: in the case of counterions only the charge density decays algebraically, while in the high salt case it decays exponentially.  

\subsection{General case}

In general, the screening effects do not write as simple exponentials and both electrostatic and osmotic contributions are complicated to assess. Eventually, one can find the exact equilibrium distance within the Poisson-Boltzmann framework \cite{Ohs75,Ben07,Pai09}. We will try to give an intuition for the result by extrapolating the above relations \eqref{counterions} and \eqref{salt2} to a more general situation. We will assume that if an equilibrium distance exists, then it should take the form of a difference between two effective counterions cloud sizes $\lambda^{\rm eff}_{\rm DNA}$ and $\lambda^{\rm eff}_{\rm prot}$ respectively brought by the DNA and the protein plates. For each plate of charge density $\sigma_X$ , this effective length has to be a function of $\lambda_X$ and $\lambda_D$. In addition, in low salt regime (i.e. $\kappa \lambda_X \ll 1$), $\lambda_X^{\rm eff} \rightarrow \lambda_X$ while at high salt concentration (i.e. $\kappa \lambda_X \gg 1$), $\lambda_X^{\rm eff} \rightarrow \lambda_X^{\rm salt}$. The only form that satisfies these constraints is:
\begin{equation}
 \lambda_X^{\rm eff} = \lambda_D \rm \: arcsinh( \kappa \lambda_X) \label{effective}
\end{equation}

A full physical analysis of this particular lengthscale in the general case of a single plate neutralized by an electrolyte can be done semi-analytically from an exact formula for the potential (see e.g. Ref. \cite{Ben07}) or numerically. Here, we will just emphasize that, after $n$ steps of size $\lambda_X^{\rm eff}$, the potential goes as $\sim \: \gamma/(\gamma + 2\kappa \lambda_X)^n$ where $\gamma > 0$ and for $n$ sufficiently big and therefore tends to zero. Depending on the value of $\kappa \lambda$, the true charge density will lie in between an algebraically decaying form and an exponentially decaying one so that the cumulative ionic charge gotten over a width $\lambda_X^{\rm eff}$ can take any value in between 50 \% and 100\% of the charge plate.

 Now, extrapolating from before we therefore assume that
\begin{equation}
 L_{eq} = | \lambda^{\rm eff}_{\rm DNA} -  \lambda^{\rm eff}_{\rm prot}| = \left|\ln \frac{\kappa \lambda_{\rm DNA}+ \sqrt{\kappa^2 \lambda_{\rm DNA}^2+1}}{\kappa \lambda_{\rm prot}+ \sqrt{\kappa^2 \lambda_{\rm prot}^2+1}} \right| \label{general}
\end{equation}This last assumption can in fact be retrieved analytically and has been tested extensively in the past \cite{Ohs75,Ben07,Pai09}.

\subsection{Energy at the minimum}

Although not intuitive, we have tried to give some motivations for the expression \eqref{general} that takes the equilibrium position at which (excess) osmotic and (screened) electrostatic pressures balance each other in the general case. Now, it so happens that the free energy per unit area at this  very equilibrium position can also be derived exactly and reads \cite{Ohs75,Pai09} 
\begin{equation}
 \beta \Delta F_{well} 
 = 4 \sigma^*\left[\sqrt{(\kappa \lambda^*)^2+1}-\kappa \lambda^*-\rm\:arcsinh \left(\frac{1}{\kappa \lambda^*}\right)\right] \,,
 \label{energy_general}
\end{equation}
where $\sigma^*$ and $\lambda^*$ are respectively the smallest surface charge density (in absolute value) and its corresponding GC length. 
In our case $\sigma^* = \sigma_{\rm prot}$.

The free energy per unit area of equation \eqref{energy_general} gives the depth of the electrostatic well at equilibrium, and is therefore a direct measure of its stability.
 Akin to Equation \eqref{general}, expression \eqref{energy_general} is quite difficult to guess, in particular because osmotic and electrostatic effects are now completely intertwined. We can still try to give a flavor of what is happening at least in the high salt regime when $\kappa \lambda_{\rm prot} \gg 1$. In this case, we make use of the fact that $\sqrt{x^2 +1} \sim x + 1/(2x) +\mathcal{O}(1/x^2)$ as $x \rightarrow \infty$ and equation \eqref{energy_general} gives thus $\beta \Delta F_{well} \sim -2 \sigma_{\rm prot}/(\kappa \lambda_{\rm prot})$. Let us try to derive this result directly, in the high salt regime. To do so, let us assume that only the screened electrostatic part $\Pi_{\rm elec} \approx -|\sigma_{\rm DNA} \sigma_{\rm prot}e^2|e^{-\kappa x}/(2 \varepsilon)$ is working and that we can neglect any osmotic effect. Integrating $\Pi_{\rm elec}$ term from infinity to $L_{eq}$ should give us an estimate of the depth of the well. We obtain
 \begin{equation}
 \Delta F_{well} \approx -\int_{\infty}^{L_{eq}}dx\:\Pi_{\rm elec}(x) \approx -\frac{|\sigma_{\rm DNA}|\sigma_{\rm prot}e^2\lambda_{\rm DNA}}{2 \varepsilon \kappa \lambda_{\rm prot}}\;\;\;\;\;\;\;{\rm (high\; salt\; regime).}
\end{equation}
Doing a little more algebra leads us to the result $ \beta\Delta F_{well} \approx -\sigma_{\rm prot}/(\kappa \lambda_{\rm prot})$ which differs from the exact formula in the high salt limit only by a factor 2 \cite{Parsegian72}. This missing factor 2 comes from the fact that there is an entropy gain from releasing salt into the bulk as the plates are brought closer from infinity and therefore, the interaction is more attractive than with screened electrostatic only \cite{Parsegian72,Ben07,PRL}. 

In the simple calculation above, we can also get some insights about why does the well depth only depends on one charge density. As we have seen, the electrostatic pressure is symmetric under the operation of exchanging the plates,  hence does not prefer one plate over the other. The equilibrium length, however, has to be positive and cares about which charge density is the smallest. This is therefore the evaluation of a symmetric term in charge densities at a position that is an asymmetric function of $\sigma$ that selects out the smallest charge density to be relevant for the energy at the minimum.

In summary, it is possible to obtain exact expressions for the position and depth of the free energy minimum corresponding to the equilibrium position induced by the balance between electrostatic attraction and osmotic repulsion (Equations \eqref{general}, and \eqref{energy_general}). These quantities depend on the plate charge densities as well as on the salt concentration \footnote{A more detailed analysis of the dependence on these quantities (and on the solution pH) can be found in Ref.s \cite{Pai09,Pai11}.}. Note moreover that Equation \eqref{energy_general}  gives a free energy {\em per unit area}, hence the total free energy is also proportional to the area of the interface.

\section{Toward a new paradigm for the target search process}
\label{paradoxsolved}


\subsection{Redefining hydrogen bonds}

Let us now come back to biology. According to our model, if the protein-DNA interface is large enough, the protein is pushed away from DNA until their distance is of the order of a fraction of nm. It is then tempting to guess that this effect can have a significant impact on the search mechanism: instead of "sticking" on DNA, proteins might "float" away from it at a very short distance, as if it were sliding on a thin cushion of air - in this case a "cushion of ions". Might its mobility along DNA  be increased? The distance between DNA and protein in the nonspecific complex allows it to slide without being hampered by the roughness associated with the sequence? And if this is the case, how may the protein still be able to distinguish the target sequence from other sequences with sufficient efficiency?

As we have discussed, recognition at the specific site is often characterized by the formation of hydrogen bonds between residues of the protein and base pairs. We have assumed that the same pattern of "possible" bonds may be used as reading frame during the search phase. 
In order to check the effect of the osmotic repulsion on this search mechanism, and therefore its balance with the specific part of the interaction, we should extend the model for this latter.
While the number of possible hydrogen bonds at each DNA position can still be described as a gaussian variable, indeed, we now need to add the energy dependence on the new problem variable: the protein-DNA distance $L$. An usual way to describe a single hydrogen bond interaction as a function of the bond length is  by a Morse potential \cite{Kang00}.
We will therefore write
\begin{equation}
\label{Morse}
 V_{\rm Morse}(L)={\mathcal E} \left[ \left(1 - \exp({-\frac{L}{\lambda_M}})\right)^2 -1 \right]
\end{equation}
where ${\mathcal E} \simeq -0.5 \,k_B T$  \cite{Tar02}  coincides with the same parameter used in the 1D model of Section \ref{Diffus}, but represents now more precisely the depth of the potential well corresponding to the bound state. In the previous expression, the parameter $\lambda_M \simeq 0.05 $ nm \cite{Che04,Nad99} is the bond range. 

Then, at each position $z=0.34\,n$ (nm) along the sequence, we suppose as before that a number ${\mathcal N}(z)$ of hydrogen bonds can be locally formed by protein with bases between $n$ and $n + N-1$. The interaction energy at position $z$ and at a distance $L$ of DNA, can be thus written as
\begin{equation}
E (z, L) = {\mathcal N}(z) \, V_{\rm Morse} (L) \,.
\end{equation}
In order to have a rather general model without referring to the case of a particular protein, we will model the  number ${\mathcal N}(z)$ of hydrogen bonds by introducing reasonable estimates of its statistical parameters and by assuming a Gaussian distribution \cite{Bar04a,Bar04b}. This assumption, as we have discussed, is in agreement with some experimental data \cite{stormo,Ger02,benichoureview12}. More precisely, we assume to know the number of bonds between the protein and its target sequence $ {\mathcal N}_{\rm max} $, which correspond to the maximum value of ${\mathcal N}$ (highest affinity). We then describe the distribution of ${\mathcal N}$ by a Gaussian with mean $ \langle {\mathcal N} \rangle = {\mathcal N}_{\rm max } / 3$ and standard deviation $\sigma_{\mathcal N} = \sqrt{{\mathcal N}_{\rm max}} $, and we furthermore impose, obviously, ${\mathcal N} \ge 0$. These values are chose so that the probability of $ {\mathcal N}= {\mathcal N}_{\rm max} $ is
realistically low. Indeed, even for sequences with a high degree of homology to the target one,
the number of H-bonds dramatically decreases, as observed e.g. in the crystal structure of non cognate BamHI complex \cite{Via00}. 

 The maximum number of bonds ${\mathcal N}_{\rm max}$ can be estimated from crystallographic data for specific complexes, and gives an average value of about 1.5 hydrogen bonds per nm$^2$ of DNA-protein interface \cite{Nad99}. For an average surface interaction $ S_{\rm prot} = 20$ nm$^2$, we obtain ${\mathcal N}_{\rm max} \simeq$ 30, and therefore   $ \langle {\mathcal N} \rangle  \simeq$10 and $\sigma_{\mathcal N} \simeq$ 5.5. With these choices, the probability of $n_{\rm max}$ bonds is reasonably low (between 3 and 4 standard deviations, Figure \ref{fig:gauss}).

\subsection{A {\em facilitated sliding}}

Summing up the two contributions, one coming  from the electrostatic interaction, the other associated with hydrogen bonds, we obtain, for the case of a protein surface charge equal to the average value found above for specific proteins (0.17 nm e$^{-2} $),  the result  presented as a free energy landscape $F(z, L)$ in figure \ref{fig:017} \cite{PRL}.

When the protein is precisely at the target, a primary
minimum exists almost at the contact with the DNA surface, corresponding to tight binding. 
Its depth is $ \sim 7 \, k_B T$ with our parameter choice.
This primary minimum is separated  by an energy barrier of the order of $k_BT$ from a secondary minimum, 
coming from the electrostatic part of the interaction. A similar scenario will be observed  in correspondence with the (rare) sequences that are close to the target sequence, and have therefore a high degree of affinity to the protein. On the contrary, for most of the positions along DNA, where the affinity is much lower, the primary minimum practically disappears and only the electrostatic equilibrium position at a distance from the DNA surface remains. 
Remarkably, the osmotic repulsion between
sequence-specific DNA-BPs and DNA dominates along non-specific
sequences : it is almost everywhere strong
enough to keep the protein at a distance from DNA, this making it in practice completely
{\em insensitive} to the sequence. 
Along the equilibrium valley, indeed, the roughness of the sequence-dependent
part of the potential is screened out: the protein can therefore
easily slide along DNA. At the target site, conversely, the large
H-bond interaction significantly reduces the barrier, and the protein
can approach the DNA.

Incidentally, the equilibrium gap distance of nearly
$0.5$~nm that we observe in Figure \ref{fig:017} is in agreement with the distance observed in the complexes
of EcoRV ($0.51$~nm \cite{Jon99}) with non-specific sequences. This also gives a rational basis to some {\it ad-hoc} protein sizes that had to be put by hand in recent coarse grained simulations of protein sliding on DNA to ensure the protein would not go closer to DNA than the distance observed in the non-specific complex \cite{Flor130,Flor131,Giv09}.

In other words, what we obtain is a mechanism that we could name {\em facilitated siding}:  the mobility of the protein is guaranteed by the osmotic repulsion, until it reaches a good
sequence and can bind it \cite{PRL}. This mechanism may represent an efficient solution of the  mobility-specificity paradox, since it introduces {\em de facto} a two-mode search: the protein is actually insensitive to the sequence all along non specific DNA, except for a few traps, and in the {\em diffusing mode} evoked in Section \ref{Diffus}. However, note that, unlike previous models, the coupling between {\em diffusing mode} and ''wrong'' sequences is here
explicit and does not require any additional ''switching'' mechanism. Moreover, in spite of the fact that the effective search obtained in our model can be intuitively described as a combination of {\em diffusing} and {\em reading mode}, the real mechanism  is in fact different: the protein is no more sensitive to the sequence, whatever the position along DNA, but it is now sensitive to  the {\em free energy barrier} that separates it from the sequence. Therefore, the interaction is always described in a similar way, but it allows for an energy activated change in the protein-DNA complex state (bringing the two bodies closer) for some special positions.
Interestingly, a similar barrier-dependent mechanism is also invoked in Ref. \cite{benichoureview12} as a solution for the mobility-specificity paradox, although the details of the model, and notably the correlation between the barrier, the primary minimum  and the sequence, are somehow different. This allows the authors to fit the available quantitative data on the search kinetics by a simple and generic kinetic model.

\subsection{Toward a different modeling of the protein search}

As we have discussed in Section \ref{models}, many theoretical models (see Ref. \cite{benichoureview12} for a good review) have been proposed to catch the essential features of the search mechanism.
We note that all these models include sliding (to different extent) and focus on dichotomic views of the search process: sliding versus 3D diffusion, "reading" versus "diffusing" modes, 
specific versus non specific binding at the target, or specific versus non specific interaction (all along the DNA).

From a numerical point of view, detailed molecular dynamics simulations seem to suggest a more complicated scenario \cite{Shoemaker00,Bou09,Chen11,PRL}  where DNA deformations, protein deformation, flexible protein tails behavior, entropic costs participate in defining a complex energy landscape for the protein-DNA complex, with rather continuous and complicated variations as a function of the the relative position of protein and DNA, either along the sequence (and therefore on and off the target) and in the radial direction, but also associated with the protein rotation and with the protein and/or DNA deformations (see \cite{Zakrzewska12} for a more exhaustive discussion).
On the other hand, it is known that a significant stabilizing effect of the specific complex is associated with the release of water molecules \cite{Bou09,Chen11}, which implies the presence of a layer of water between proteins and DNA in the nonspecific complex.

Very interestingly, the scenario obtained by our model shares some central features with what is found numerically by some authors. In particular, either Ref. \cite{Bou09} and 
\cite{Chen11} evidence the presence of two distinct free energy minima, one closer to DNA, the other farer from it, separated by a free energy barrier. The relative positions of the three states are smaller but not incompatible with what obtained in our model \footnote{In Ref. \cite{Bou09}, the secondary minimum, barrier and primary minimum locations are found respectively at protein-DNA distances of 0.32, 0.31 and $<$0.3 nm. In Ref. \cite{Chen11}, at 0.26, 0.13 and 0.08 nm, respectively.}.

These finding suggest an alternative way of describe the search process, by replacing the usual dichotomic view by a more "soft" approach where the interaction is described in terms of {\em continuous variables}. 
The protein-DNA distance is indeed a crucial variable, potentially leading to a description of the protein kinetics where the distinction between {\em sliding}, {\em hopping}, {\em jumping} and 3D diffusion becomes somehow obsolete. 
More concretely, the movement of the protein in the vicinity of DNA can, in our scenario, be treated as a diffusion in the landscape of figure \ref {fig:017}. 
Unfortunately,  {\em in vitro} experiments, which assess for sliding cannot reach the resolution needed to describe the protein DNA interaction (and associated kinetics) at the scale involved in this model. However, experimentalists clearly distinguish at least phases where the proteins are ''on'' DNA (and can therefore be observed) from phases where the protein dissociates from it. Moreover, rapid displacements along a same DNA molecule have been observed \cite{Bonnet08} that cannot be compatible with pure 1D diffusion along the double helix. The question therefore arise of how these different protein {\em states} or {\em modes of displacement} can be accounted for in the context of a continuous description.

\subsection{Defining a physical-meaningful {\em sliding} time}

By comparing our model to experimental estimates of the chemical rates of protein binding and unbinding, one can in principle get more decisive feedback
about the landscape, since binding and unbinding events involve a wide range of
DNA-protein distances. In the following, for the sake of simplicity, we shall
focus on the dissociation rate of a non-specific protein-DNA complex 
although the binding rate can also be considered without too much difficulty
following e.g. Ref. \cite{Berne88}. Moreover, we neglect here the effects due to the hydrogen bond interaction, only relevant at very short distances : the aim of this calculation is indeed to evaluate the time needed for the protein to escape from a generic, non specific position along DNA, i.e. to exit the secondary minimum defined by the electrostatic part of the interaction.

We are interested in the following reaction:
\begin{equation}
\rm (Prot|DNA)_{complex} \: \rightarrow \:\rm Prot + DNA \label{dissoc}
\end{equation} 
We will assume that the size of the particles is big enough for the unbinding process to be diffusion dominated \cite{Berne88}. Considering the energy landscape we derived in the previous parts, it is natural to use the surface-to-surface DNA-protein separation $L$ as the reaction coordinate. Moreover, if the energy landscape displays a well defined barrier between the two chemical states of reaction (\ref{dissoc}), then we can use Kramers' rate theory for a one-dimensional isomerization process \cite{Berne88}. The dissociation rate $k_{diss}$ reads then:
\begin{equation}
k_{diss} \approx \frac{D}{2 \pi} \sqrt{ \beta  |G''(L_A)||\beta  G''(L_B)|} \:e^{-\beta \Delta_{AB} G} \label{Kramers}
\end{equation}where $D$ is the diffusion constant of the protein, $A$ corresponds to the minimum of the binding well, $B$ is the location of the dividing surface i.e. the top of the energy barrier (cf. Fig. \ref{Paysage_radial}), $\Delta_{AB} G=G_B-G_A$  and $G''$ stands for a second derivative of the energy $G$ with respect to $L$. The total effective interaction $G(L)$ in Eq. (\ref{Kramers}) is defined so that the ratio of the marginal probabilities to be either at $L_1$ or $L_2$ reads:
\begin{equation}
\frac{p(L_1)}{p(L_2)} \equiv \frac{e^{-\beta G(L_1)}}{e^{-\beta G(L_2)}} \label{boltzmann1}. 
\end{equation}
for any $L_1$ and $L_2$. 

On the other hand, it is also possible to state that this same ratio  should read:
\begin{equation}
\frac{p(L_1)}{p(L_2)} \equiv \frac{2\pi(R_{DNA}+ L_1) e^{-\beta F(L_1)}}{2\pi (R_{DNA}+L_2) e^{-\beta F(L_2)}} \label{boltzmann2}
\end{equation}
where the $F(L)$ is the free energy (that we estimated in previous Sections) that corresponds to the work one has to do to bring a protein from infinity to a distance $L$ from a DNA segment for any fixed value of the polar angle that locates the protein within the plane perpendicular to the DNA axis. 
The $2\pi (R_{DNA}+L)$ factor is a degeneracy term, associated to the probability of being at a particular distance from the axis of the DNA molecule. This probability grows indeed as the circumference of a circle of radius $R_{DNA}+L$. 

Note that, unlike $F(L)$, $G(L)$ may present a maximum , i.e. an energy barrier between the location of the electrostatic minimum and the region $L\to \infty$ (see Figure \ref{Paysage_radial}). 
From Eqs. (\ref{boltzmann1}) and (\ref{boltzmann2}), we thus find that the total effective interaction $G(L)$ associated to a distance $L$ has to have the form:
\begin{equation}
G(L) = F(L) - k_B T\ln\left(\frac{R_{DNA}+L}{R_0}\right) \label{total_free_energy}
\end{equation}
where $R_0$ is some unimportant distance whose purpose is to have a dimensionless argument inside the logarithm. Now that we have understood that, we can try to interpret Kramers' formula~(\ref{Kramers}). To do so, we rewrite (\ref{Kramers}) in a slightly different way:
\begin{equation}
k_{diss} \approx \frac{1}{2 \pi} \sqrt{\frac{D}{\delta L_A^2}\frac{D}{\delta L_B^2}} e^{-\beta \Delta_{AB} G} = \frac{\sqrt{\nu_A\nu_B}}{2\pi} e^{-\beta \Delta_{AB} G} \label{Kramers_new}
\end{equation}where $\delta L_A \equiv 1/\sqrt{\beta G''(L_A)}$ and $\delta L_B \equiv 1/\sqrt{\beta G''(L_B)}$ are respectively the typical sizes of the bottom of the well and the top of the barrier and where $\nu_A^{-1} \equiv \delta L_A^2/D$ and $\nu_B^{-1} \equiv \delta L_B^2/D$ are the typical times it takes for a diffusive protein to travel over the lengths $\delta L_A$ and $\delta L_B$ respectively. Thus, the pre-factor $\sqrt{\nu_A \nu_B}$ is nothing but the geometric mean of the natural rates $\nu_A$ and $\nu_B$.
To get some insights from Eq.~\eqref{Kramers_new}, we first calculated $k_{diss}$ from the model with parameters used for Figure \ref{Paysage_radial}, i.e. in the case of a physiological salt concentration $n_b = 0.1$mol L$^{-1}$. 
We found that $\beta \Delta_{AB}G \approx 3.4$ while the pre-factor $\sqrt{\nu_A \nu_B}/2\pi \approx 10^3\:\rm ms^{-1}$. Overall the rate is $k_{diss}[0.1\:M] \approx 35\:\rm ms^{-1}$. It thus means that on average in physiological conditions a protein with a landscape as that of Fig. \ref{Paysage_radial} will stay less than a millisecond on a given DNA segment before leaving it.
This observation seems however in contradiction with measured average sliding times in experiments \cite{Bonnet08} where a protein can be bound to a DNA segment for up to few seconds. This discrepancy is without accounting for the fact that the mentioned experiments are done at much lower salt concentration. In fact, as we have seen before, increasing the salt concentration can have a very strong effect on the free energy landscape. We thus recalculated it with the same protein and DNA parameters but with $n_b = 0.01\:M$. We got that $\beta \Delta_{AB}G \approx 9$ while the pre-factor in Eq. \eqref{Kramers_new} is about $10^2\:\rm ms^{-1}$. Overall the dissociation rate $k_{diss}$ is $k_{diss}[0.01\:M]\approx 10^{-2} \:\rm ms^{-1}$ which is about three orders of magnitude lower than in physiological conditions! Also, in this particular case, the typical life time of the non-specific complex is comparable to those observed in Ref. \cite{Bonnet08}.

In this part, we were able to relate our continuous description to observable quantities such as the dissociation rates of the non-specific complex of arbitrary proteins. To apply Kramers theory, we emphasized the fact that the reaction coordinate is a radial coordinate that gives rise to an entropic repulsive force that allows for a non ambiguous definition of the barrier between the bound state and the unbound one. In absence of the mentioned $2\pi(R_{DNA}+L)$ degeneracy however (i.e. in a truly one dimensional case), there is no consensus on where to put the dividing surface for free energies as those of Fig. \ref{fig:twoplates} and one should be careful about this point \cite{Kramers}.

Evidently, the next step in exploiting the model described here will be to try to predict the features of the protein diffusion along DNA during sliding, and to compare them with experiments.
Note however that, although we can in principle estimate the sliding diffusion coefficient $D_1$ from diffusion properties of the protein in bulk and get an estimate of the typical sliding length ($\sim \:\sqrt{D_1/k_{diss}}$) that is measured in many experiments ({\em in vitro} but also {\em in vivo}, see e.g. \cite{Elf12}), it is in fact more subtle than expected. Indeed, as it was imagined by Schurr \cite{Sch79}, some DNA-binding proteins  slide with an helical motion along DNA \cite{Blainey09,Dikic12}. The resulting effective diffusion coefficient then depends on the DNA-protein distance in the bound state \cite{Bla08,Blainey09, Dikic12} and we have seen that the latter depends on the salt concentration; the sliding diffusion coefficient therefore depends on the salt concentration.
This additionally supports a potential need for the change of paradigm that has been stressed throughout this chapter in order to understand fully what are the relevant parameters to describe the observed binding kinetics of proteins to their specific sites on DNA.


\newpage

\bibliographystyle{unsrt} 
\bibliography{bibli} 

\newpage

\section*{Figure legend\\ \small (a small reproduction of the figure is added for clearness)}

\begin{figure} [h]
{\centerline{\includegraphics [width = .7 \textwidth]{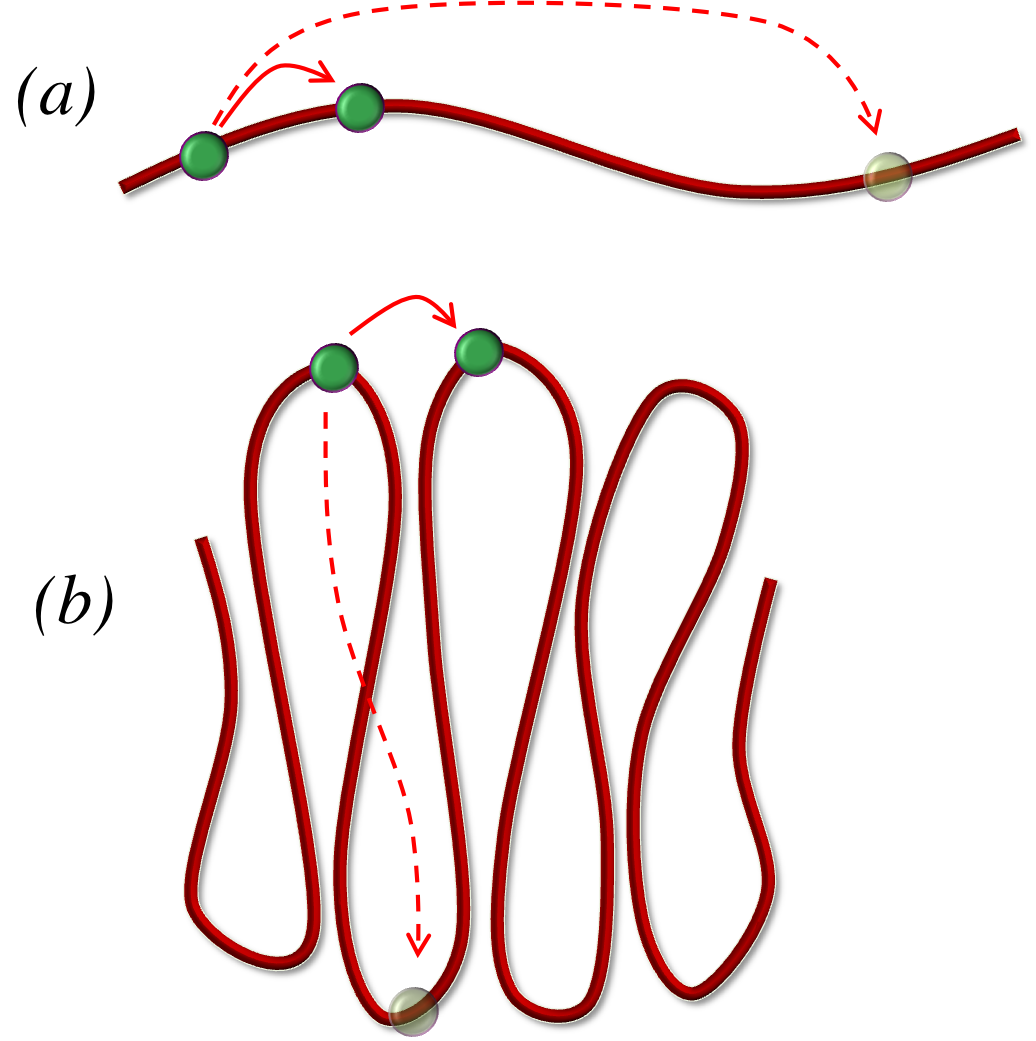}}}
\caption{In the scenario proposed by Kupiec, the diffusion of proteins is responsible for the activation of different genes. The distance of these genes at the position of the site where the transcription factor is synthesized determines the speed of search and therefore the efficiency of activation, either in the case when the linear distance along the molecule is concerned {\em (a)}, or the three-dimensional distance due to the arrangement of the DNA into the nucleus {\em (b)}. }
\label{fig:kupiec}
\end{figure}

\begin{figure} [h]
{\centerline{\includegraphics [width = .7 \textwidth]{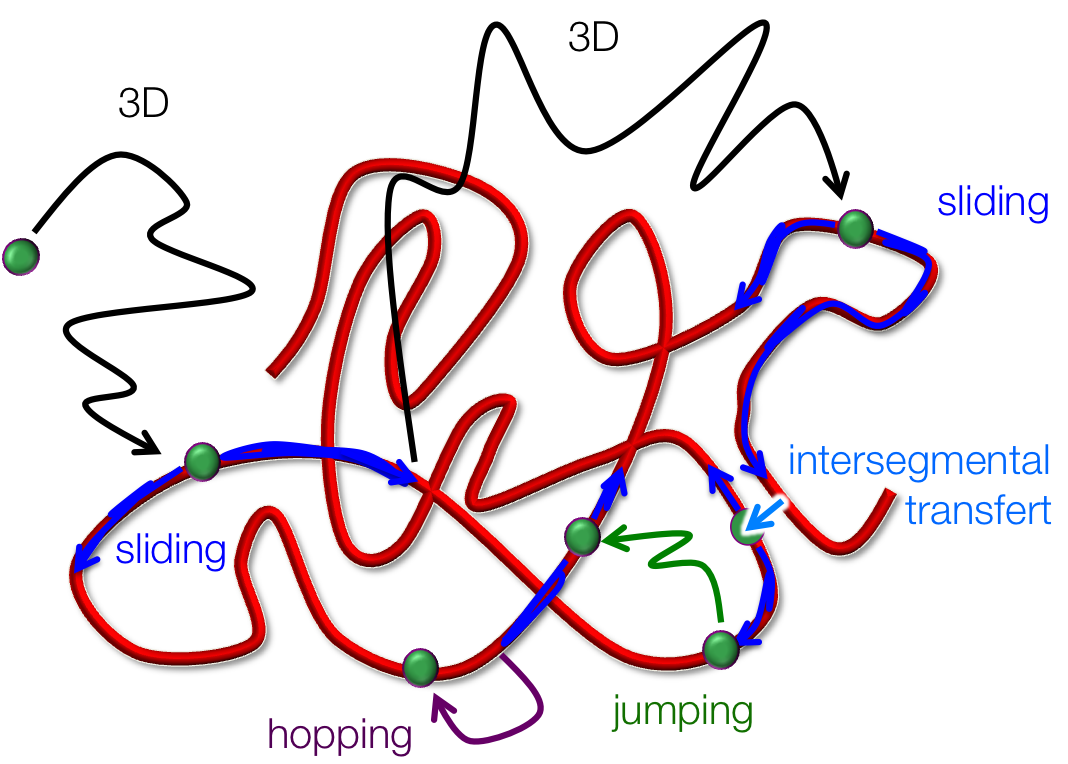}}}
\caption{The search modes usually considered in literature: 3D diffusion,  {\em sliding} or 1D diffusion along the double helix, {\em hopping} at a close site, {\em jumping} to a different DNA stretch, and {\em intersegmental transfer}, involving simultaneous binding to two distinct DNA stretches. }
\label{fig:hopping}
\end{figure}

\begin{figure} [!h]
\begin{center}
{\centerline{\includegraphics*[width = 0.7 \textwidth]{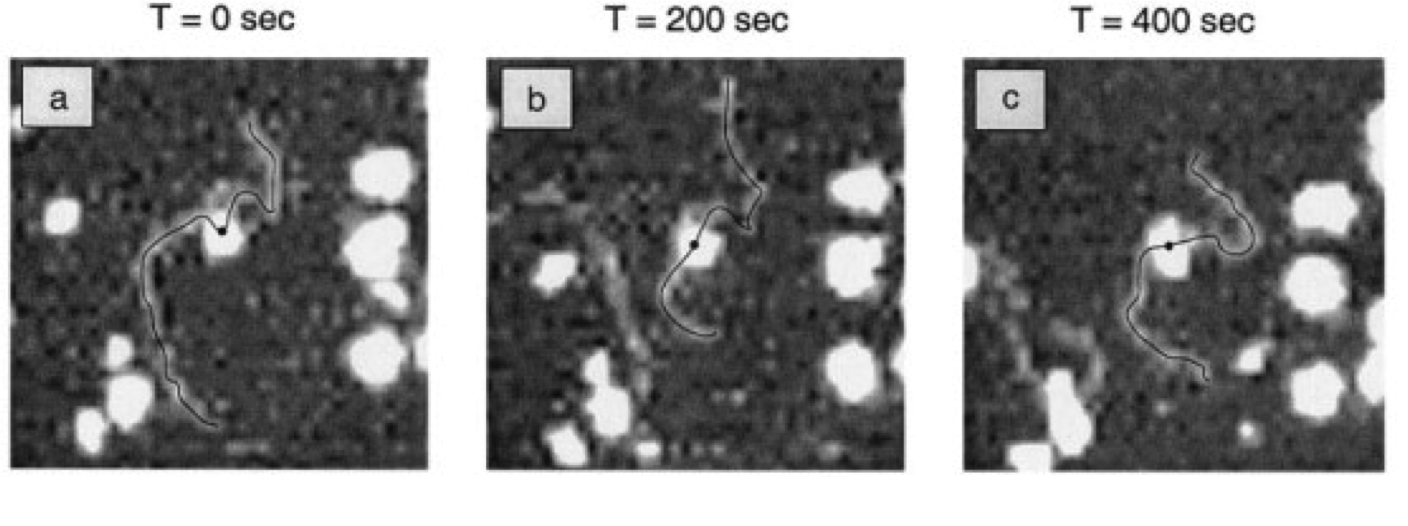}}}
\caption{Three successive AFM images showing the complex formed by the RNA polymerase of E. Coli, fixed on a mica surface, and a non-specific DNA sequence, semi-adsorbed on the same surface. This type of experience can show the
relative movement of the protein along DNA, but fails in giving a quantitative description of the diffusion due to geometrical constraints. Figure adapted from Ref. \cite{Gut99}.}
\label{AFM}
\end{center}
\end{figure}

\begin{figure} [h!]
\begin{center}
{\centerline{\includegraphics*[width = 0.7 \textwidth]{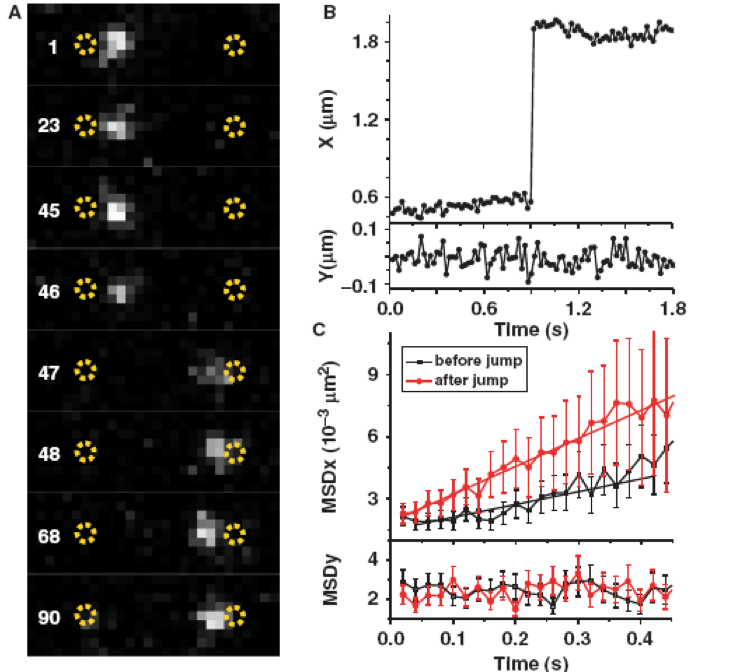}}}
\caption{Figure of Ref. \cite{Bonnet08} in which {\em sliding}  and {\em jumping} events are directly observed. {\bf A}: 
Subsequent fluorescent images of the protein (white spot) moving along a stretched DNA (yellow circles on both sides of the figure shows the two ends of the DNA segment).  
Between frames 46 and 47,  a {\em jump} can be observed. {\bf B} Longitudinal ($X$) and transverse ($Y$) 
displacement of the protein as a function of time. 
The jump of about 1300 nm is again detected in the X-trajectory. 
{\bf C}: The longitudinal MSD calculated before and after the jump display 1D diffusion similar to that observed during events without large jumps. Values of the diffusion constant are between  0.3 and 0.6  10$^{?2}$ $\mu$m$^2$/s. 
(Isabelle Bonnet, Andreas Biebricher, Pierre-Louis Porté et al. Sliding and jumping of single EcoRV restriction enzymes on non-cognate DNA. Nucl. Acids Res. (2008) 36(12): 4118-4127, Figure 3. By permission of Oxford University Press).}
\label{cap}
\end{center}
\end{figure}

\begin{figure} [h!]
\begin{center}
{\centerline{\includegraphics*[width = 0.7 \textwidth]{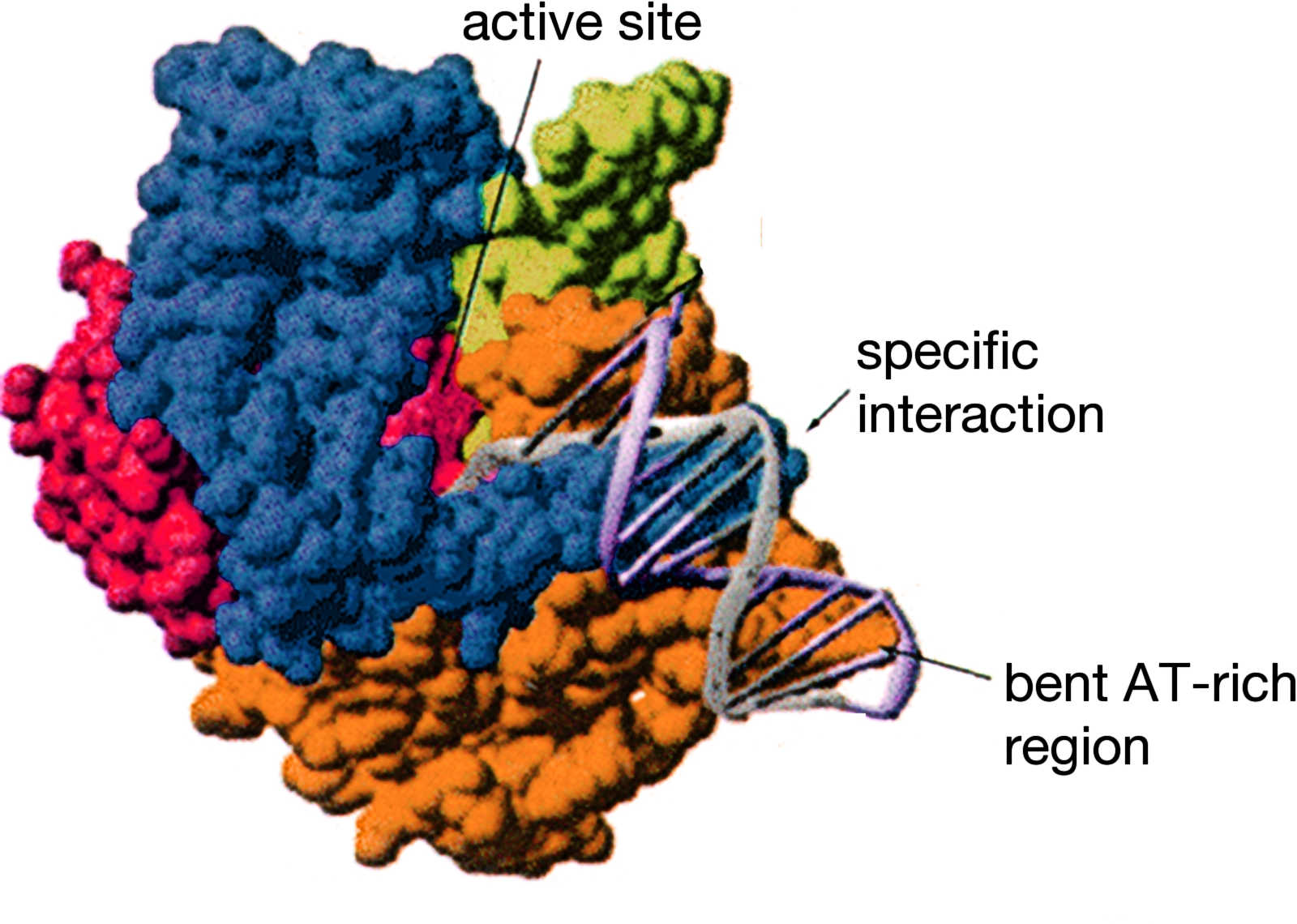} }}
\caption{Crystallographic reconstruction of the interaction between the RNA-polymerase and its T7 target sequence. The three interaction regions mentioned in the main text are indicated. Adapted from Ref. \cite{Cheetham99}.}
\label{fig:T7}
\end{center}
\end{figure}

\begin{figure} [!h]
\begin{center}
{\centerline{\includegraphics*[width = 0.7 \textwidth]{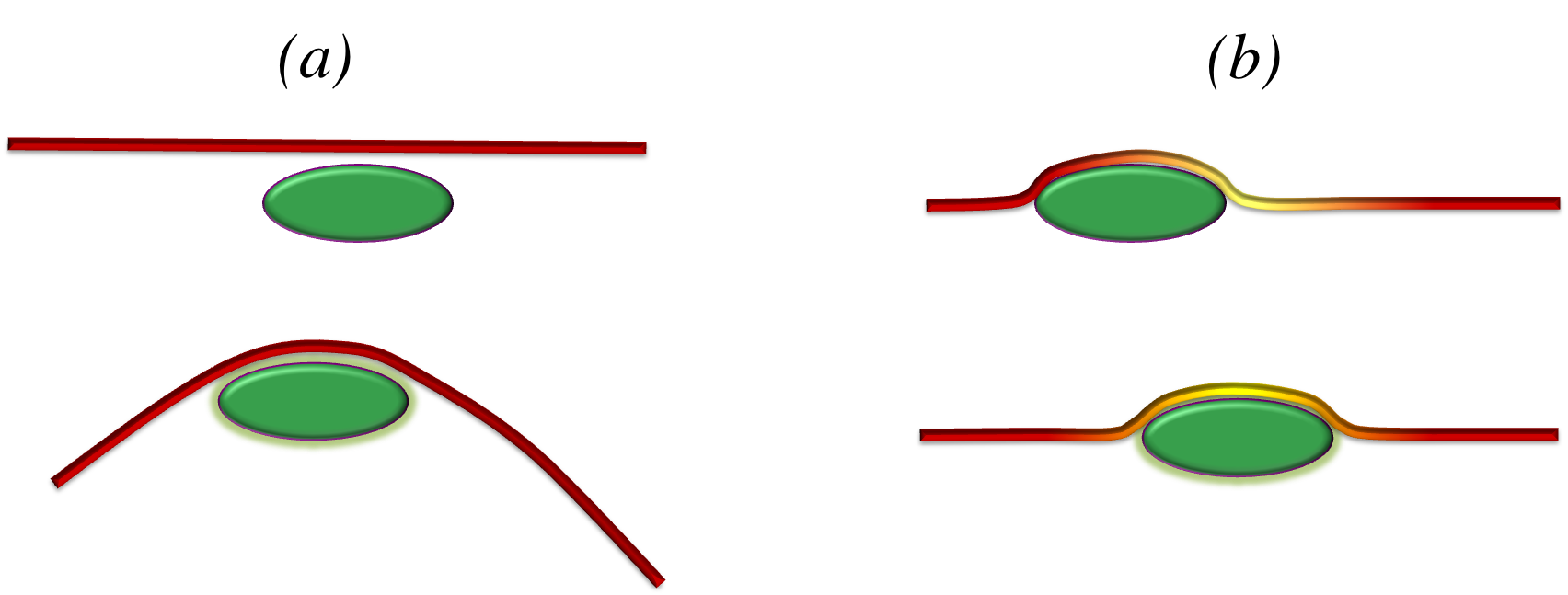}}}
\caption{Local DNA curvature ({\em a}) or flexibility (yellow region, {\em b}) can affect the protein-DNA interaction. This physical properties being sequence-dependent, this provides a sequence-dependent contribution to the interaction energy profile. With respect to direct chemical bond, the curvature/flexibility effect is expected to vary in a smoother fashion.
}
\label{fig:slusky2}
\end{center}
\end{figure}

\begin{figure} [!h]
\begin{center}
{\centerline{\includegraphics*[width = 0.7 \textwidth]{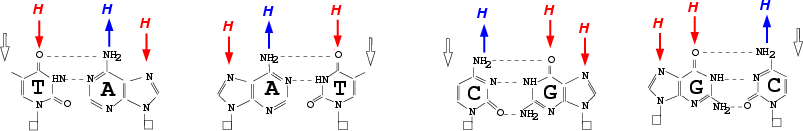}}}
\caption{Hydrogen bond acceptor (red) and donor (blue) sites on the four base pairs accessible through the major groove. Note that a similar four-sites pattern can be defined for each base pair, but associated with a different acceptor/donor order. 
}
\label{fig:Hbonds}
\end{center}
\end{figure}

\begin{figure} [!h]
\begin{center}
{\centerline{\includegraphics*[width = 0.7 \textwidth]{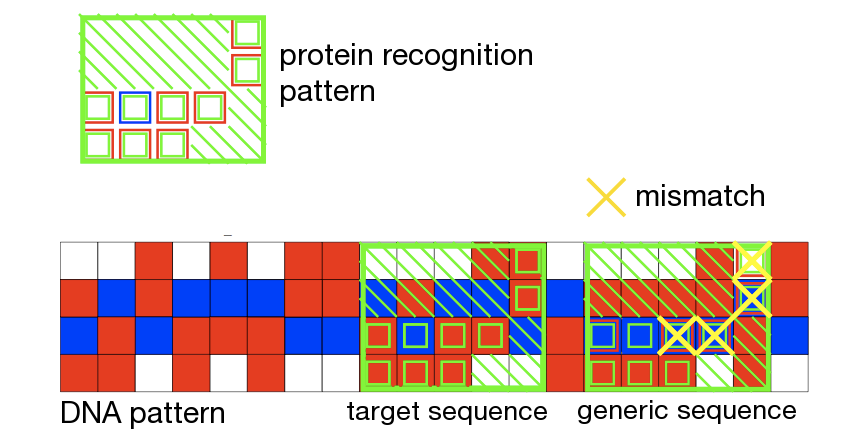}}}
\caption{While sliding along DNA, the protein applies a recognition pattern to {\em read} the sequence by counting the number of acceptor or donor groups that corresponds to its own motif.
}
\label{fig:mismatch}
\end{center}
\end{figure}

\begin{figure} [!h]
\begin{center}
{\centerline{\includegraphics*[width = 0.7\textwidth]{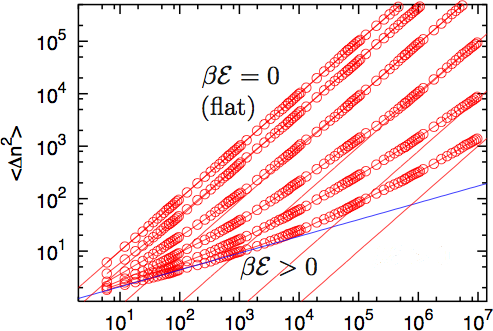}}}
\caption{Mean squared displacement obtained by simulating the diffusion of a particle on the rough energy profile associated with by hydrogen bonding and defined in the main text.  From the upper curve to the bottom: $\beta {{\mathcal E}} = 0,\,0.3,\,0.6,\,0.9,\,1.2,\,1.5$. Red lines of slope 1 and one blue line of slope $0.3$ are reported for comparison.
}
\label{fig:subdiff}
\end{center}
\end{figure}

\begin{figure} [h!]
\begin{center}
{\centerline{\includegraphics*[width = 0.7\textwidth]{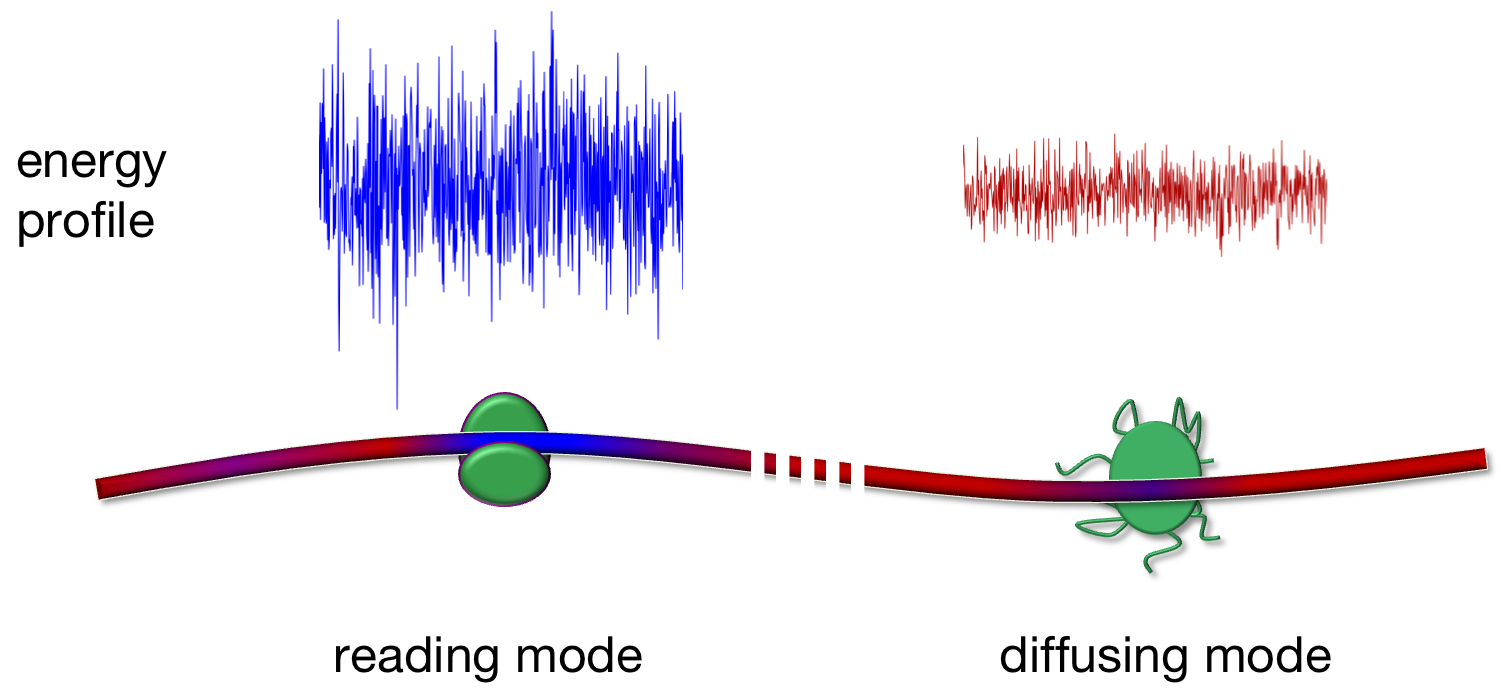}}}
\caption{Slusky \& Mirny hypothesize that partial denaturation of the protein may be responsible for a significant change in the {\em effective} energy profile associated with the interaction with DNA. In the {\em diffusing mode}, the partially denatured protein is much less sensitive to the sequence and its mobility is therefore increased \cite{Slu04}.}
\label{fig:Slusky}
\end{center}
\end{figure}

\begin{figure} [h]
\begin{center}
{\centerline{\includegraphics*[width = 0.7\textwidth]{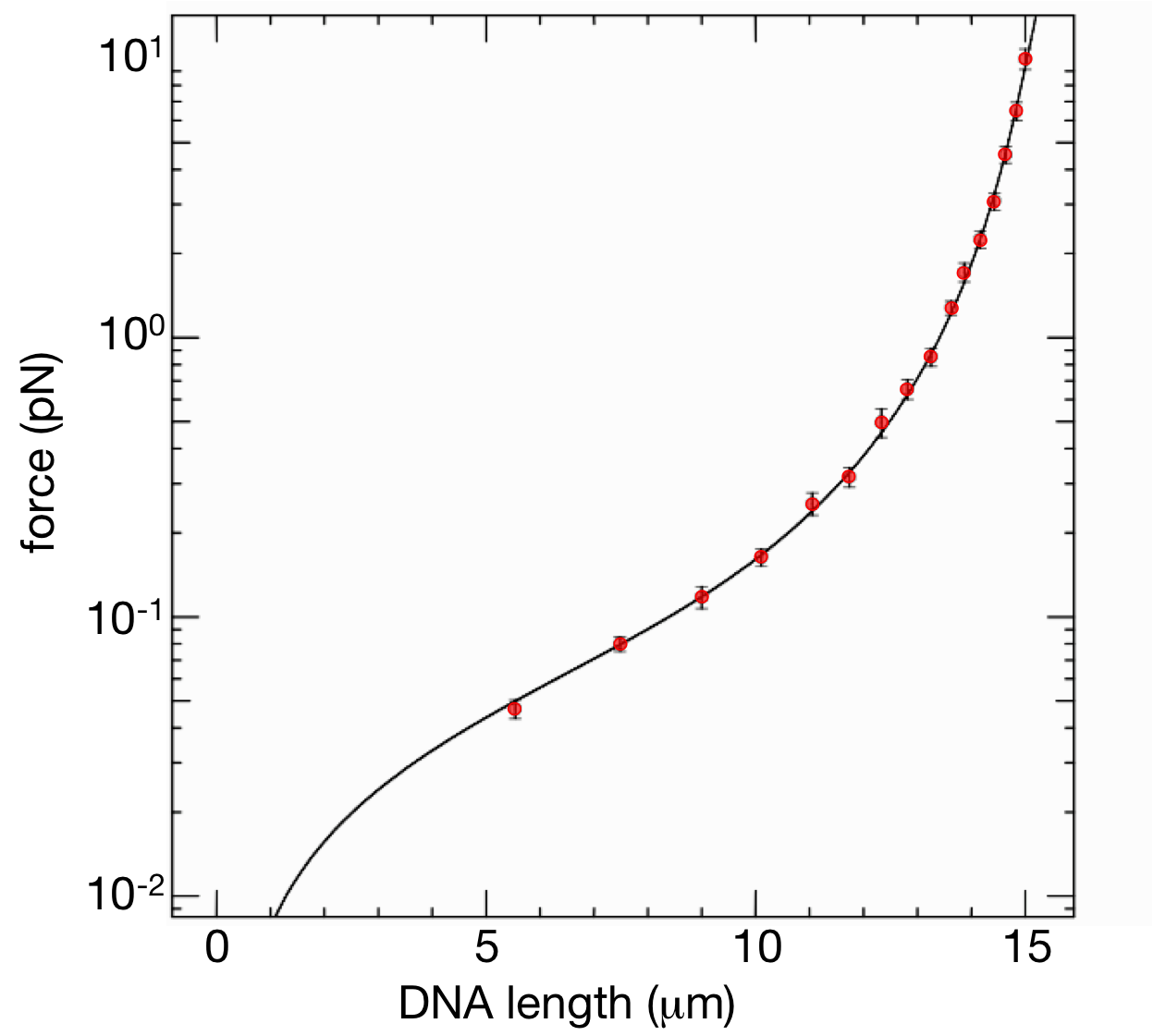} }}
\caption{\label{fig3}  Typical experimental results for the extension of DNA when subjected to a constant force, fitted by the {\em Worm Like Chain} model.}
\end{center}
\end{figure}

\begin{figure} [!h]
\begin{center}
{\centerline{\includegraphics[width = 0.7\textwidth]{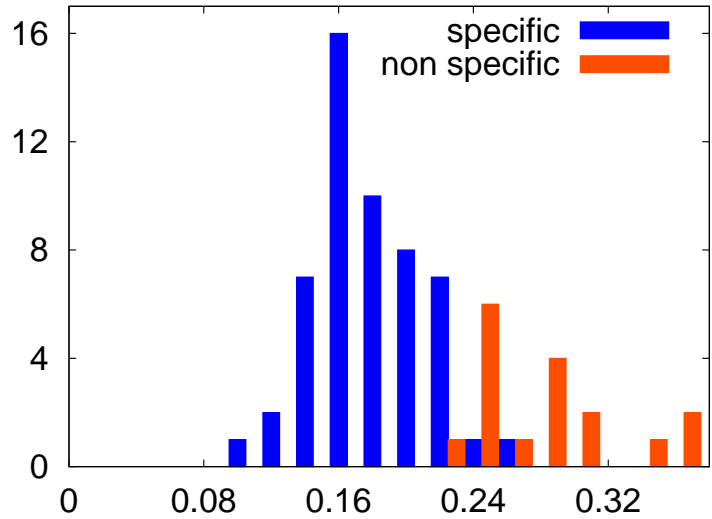}}}
\caption {Histogram of the surface charge density of the interface binding proteins to DNA. Specific (blue) and non specific (orange) proteins are separately considered.}
\label{fig:history}
\end{center}
\end{figure}

\begin{figure} [!h]
\begin{center}
{\centerline{\includegraphics*[width = 0.7\textwidth]{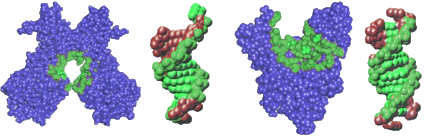} }}
\caption{Two example of complementary-shape proteins, adapted from Ref. \cite{Jon99}.  Left : NF-kB (1nfk); right: EcoRI restriction endonuclease (1eri). In blue are represented residues of the protein that do not contact DNA (in red).  All protein and DNA groups which come in close contact and form the interface in the protein-DNA complex are shown in green.}
\label{fig:jones}
\end{center}
\end{figure}

\begin{figure} [ht!]
\begin{center}
{\centerline{\includegraphics[width = .7\textwidth]{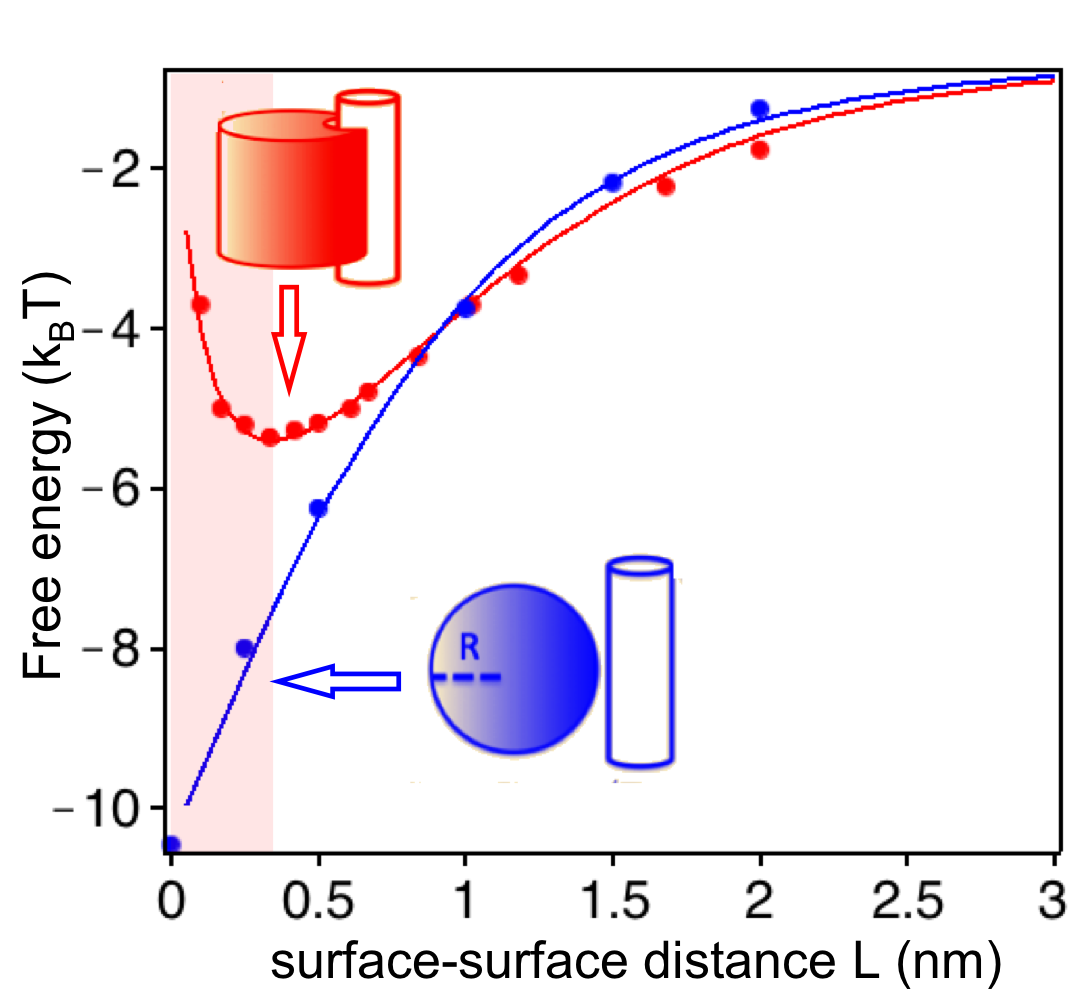}}}
\caption{Monte Carlo (points) and Poisson-Boltzmann (lines) estimations of the protein-DNA electrostatic interaction for two different protein shapes : a spherical one (blue) and concave, DNA-matching one (blue). In both cases, the results from Poisson-Boltzmann theory applied to the two-plates geometry are adapted to the curved surfaces by mean of a Derjaguin approximation. In the concave case clearly the osmotic repulsion is clearly observed, while it is absent in the spherical case due to the highly limited area of the interface. }
\label{fig1}
\end{center}
\end{figure}

\begin{figure} [ht!]
\begin{center}
{\centerline{\includegraphics*[width = .7\textwidth]{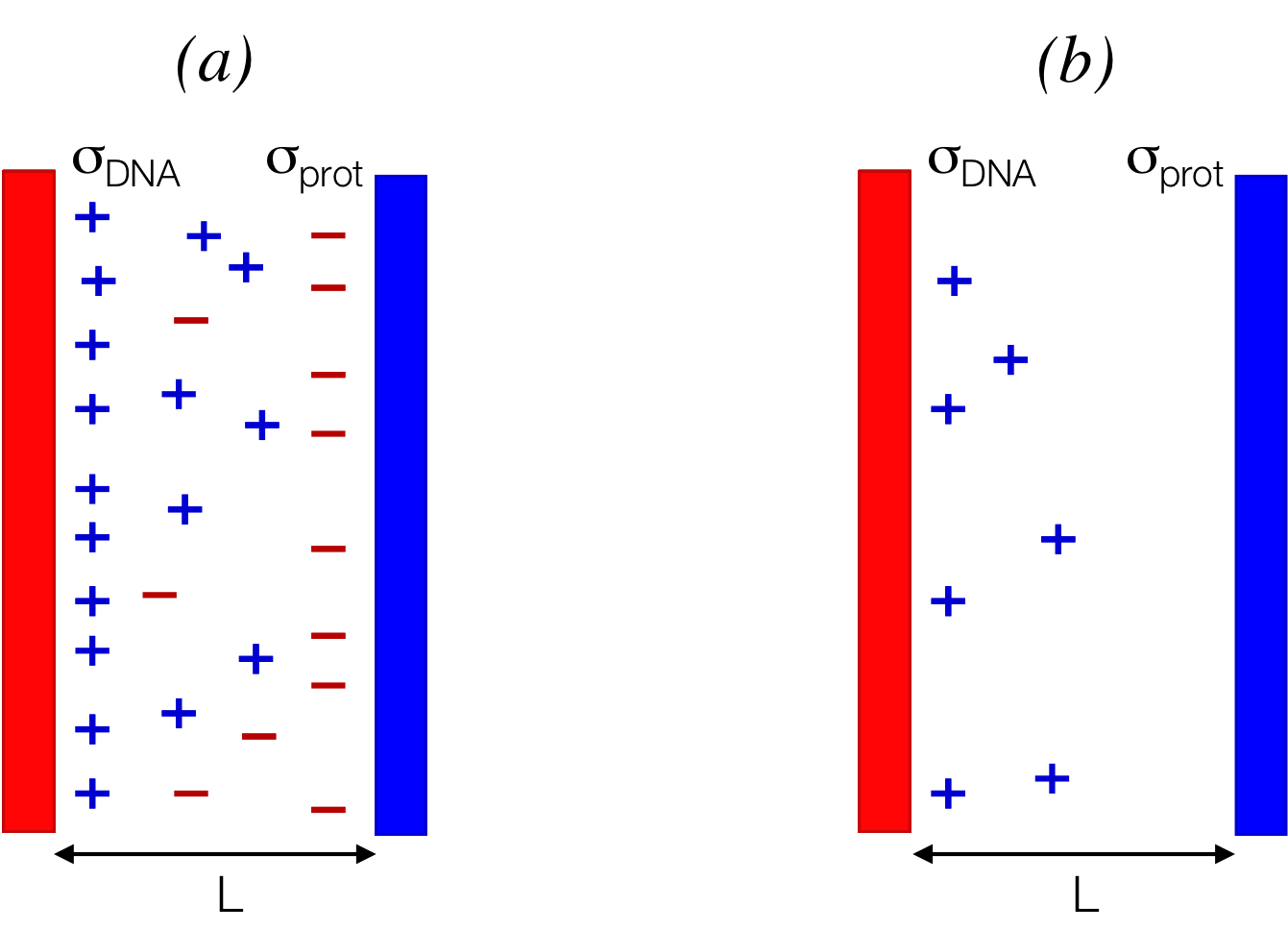}}}
\caption{The two plates system discussed in the theoretical section, both in presence of salt {\em (a)} or in the counterions only regime {\em (b)}. }
\label{fig:twoplates}
\end{center}
\end{figure}

\begin{figure}[!h]
\begin{center}
{\centerline{\includegraphics*[width=0.7\textwidth]{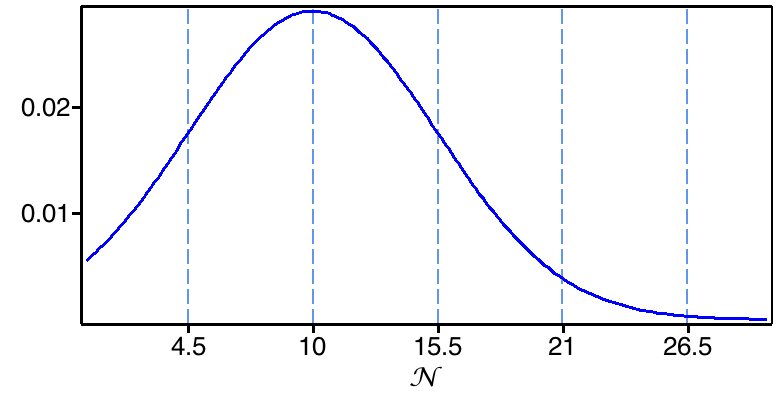}}}
\caption{Gaussian distribution of the parameter $\mathcal N$, corresponding to the number of possible hydrogen bonds between the protein and the DNA, within 0 and ${\mathcal N}_{\rm max}$, using  the parameters defined in the main text. The vertical dashed lines are centered on the mean value  and are separated by one standard deviation.}
\label{fig:gauss}
\end{center}
\end{figure}

\begin{figure}[!h]
\begin{center}
{\centerline{\includegraphics*[width=0.7\textwidth]{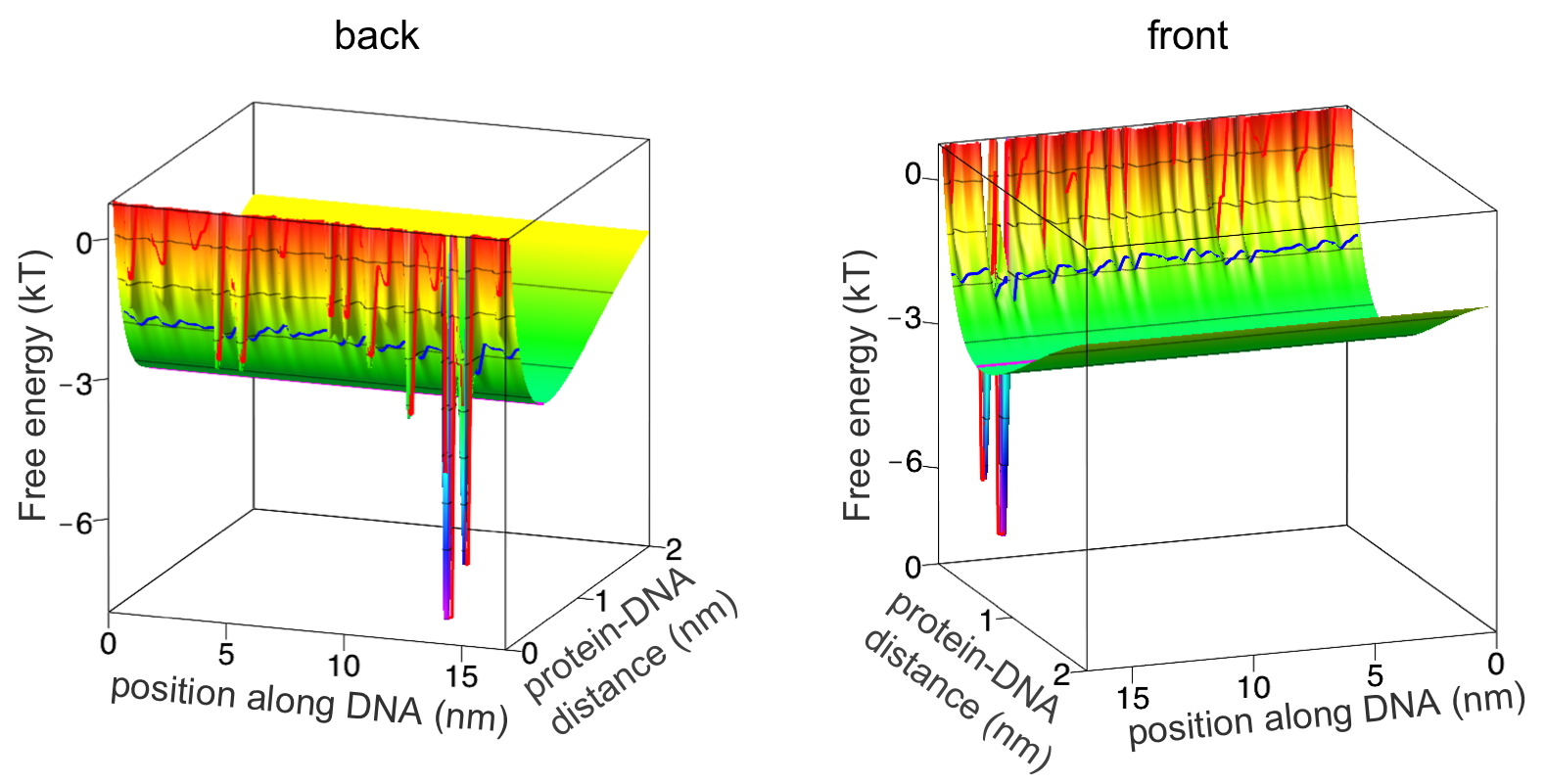}}}
\caption{
Free energy is here calculated along a DNA sequence of 50 bp, as a function of the protein-DNA distance $ L $ and of the position $ z $ of the protein along the DNA, for $ \sigma_ {\rm prot} = 0.17 \, e$ nm $^{-2} $. The distance between the contour lines is $k_BT$. For clarity, we show the same graph from two opposite sides (back and front). A red and a blue curves are added as a guide for the aye in the approximate position (along DNA) of the primary minimum and of the barrier, respectively.}
\label{fig:017}
\end{center}
\end{figure}

\begin{figure}[!h]
\begin{center}
{\centerline{\includegraphics*[width=0.7\textwidth]{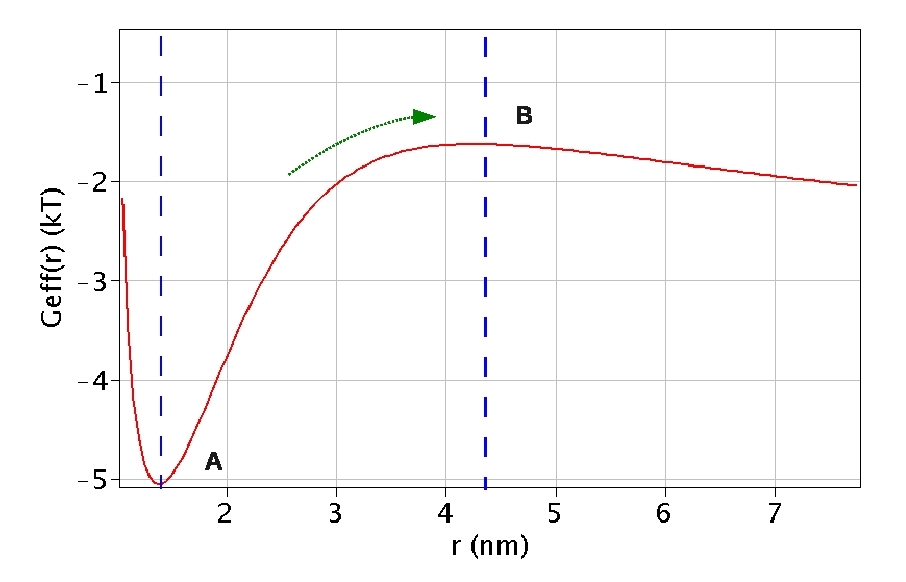}}}
\caption{\label{Paysage_radial}Free energy landscape for the radial coordinate. The curve represents the thermodynamic potential $G$ associated with the effective diffusion in the radial direction under physiological conditions (i.e., $ c_ {\ rm salt} = 0.1$mol L$^{-1}$). Point A corresponds to the bound state allowing a one dimensional diffusion along DNA while point B is the point beyond which the protein can diffuse freely in three dimensions and therefore corresponds to the dividing surface. }
\end{center}
\end{figure}

\end{document}